\documentclass[12pt]{article}
\usepackage{graphicx}
\usepackage[numbers]{natbib}
\usepackage{sidecap}
\usepackage{wasysym}
\usepackage{hyperref}

\newcommand{\mic}{$\mu$m}
\newcommand{\apj}{ApJ}
\newcommand{\apjl}{ApJL}

\newcommand{\nat}{Nature}
\newcommand{\mnras}{MNRAS}

\newcommand{\prd}{Phys. Rev. D}
\newcommand{\aap}{Astron. Astrophys.}
\newcommand{\aj}{Astron. J}

\usepackage{color}
\usepackage{times}
\usepackage{geometry}
\usepackage[utf8]{inputenc}
\usepackage{enumitem,amssymb}
\usepackage{ragged2e}
\newlist{thematic}{itemize}{8}
\setlist[thematic]{label=$\square$}
\usepackage{pifont}

\begin{document}
\raggedright
\huge
Astro2020 Science White Paper \linebreak

Electromagnetic probes of primordial black holes as dark matter  \linebreak
\LARGE
\normalsize

\noindent \textbf{Thematic Areas:} \hspace*{60pt} $\square$ Planetary Systems \hspace*{10pt} $\square$ Star and Planet Formation \hspace*{20pt}\linebreak
$\XBox$ Formation and Evolution of Compact Objects \hspace*{31pt} $\XBox$ Cosmology and Fundamental Physics \linebreak
  $\square$  Stars and Stellar Evolution \hspace*{1pt} $\square$ Resolved Stellar Populations and their Environments \hspace*{40pt} \linebreak
  $\square$    Galaxy Evolution   \hspace*{45pt} $\XBox$             Multi-Messenger Astronomy and Astrophysics \hspace*{65pt} \linebreak
  
\textbf{Principal Author:}

Name:	A. Kashlinsky
 \linebreak						
Institution:  Code 665, Observational Cosmology Lab, NASA Goddard Space Flight Center, Greenbelt, MD 20771 and SSAI, Lanham, MD 20770
 \linebreak
Email: Alexander.Kashlinsky@nasa.gov
 \linebreak
Phone:  301-286-2176
 \linebreak
 
\textbf{Co-authors:} 
Y. Ali-Ha\"imoud (NYU), 
S. Clesse (U. Louvain/Namur U.), 
J. Garcia-Bellido (IFT-UAM/CSIC Madrid), 
L. Wyrzykowski (Warsaw U.) +  
A. Achucarro (U. Leiden), 
L. Amendola (U. Heidelberg), 
J. Annis (FNAL)
% ...
\linebreak
 \linebreak
\justifying
\noindent
\textbf{Abstract:}
The LIGO discoveries have rekindled suggestions that primordial black holes (BHs) {\it may constitute part to all of the dark matter} (DM) in the Universe. Such suggestions came from 1) the observed merger rate of the BHs, 2) their unusual masses, 3) their low/zero spins, and 4) also from the independently uncovered cosmic infrared background (CIB) fluctuations signal of high amplitude and coherence with unresolved cosmic X-ray background (CXB). Here we summarize the prospects to resolve this important issue with electromagnetic observations using the instruments and tools expected in the 2020's. These prospects appear promising to make significant, and potentially critical, advances. We demonstrate that {\bf in the next decade, new space- and ground-borne  electromagnetic instruments, combined with concurrent theoretical efforts, should shed critical light on the long-considered link between primordial BHs and DM.}
Specifically the new data and methodologies under this program will involve:
\begin{itemize}
\item Probing with high precision the spatial spectrum of source-subtracted CIB with {\it Euclid} and {\it WFIRST}, and its coherence with unresolved cosmic X-ray background using {\it eROSITA} and {\it Athena},\vspace{-3mm}
\item Advanced searches for microlensing of Galactic stars by the intervening Galactic Halo BHs with OGLE, {\it Gaia}, LSST and {\it WFIRST},\vspace{-3mm}
\item Supernovae (SNe) lensing in the upcoming surveys with {\it WFIRST}, LSST and also potentially with {\it Euclid} and {\it JWST},\vspace{-3mm}
\item Advanced theoretical work to understand the details of PBH accretion and evolution and their influence on cosmic microwave background (CMB) anisotropies in light of the next generation CMB experiments,\vspace{-3mm}
\item Better new samples and theoretical understanding involving stability and properties of ultra faint dwarf galaxies, pulsar timing, and cosmological quasar lensing.
\end{itemize}
\pagebreak
\clearpage
The LIGO discovery of the first gravitational wave (GW) signal from two merging black holes (BHs) \citep{Abbott:2016b,Abbott:2016c} rekindled suggestions that primordial BHs (PBHs) \citep{Hawking:1974sw,Carr:1974nx,1975Natur.253..251C} constitute most or all of the dark matter (DM). The suggestions were based on the deduced rate of their mergers \citep{Clesse:2017,Bird:2016} as well as the properties  of the near-IR cosmic infrared background (CIB) uncovered in 2005 \citep{Kashlinsky:2016}. Latest LIGO O1+O2 results find 10 significant BH mergers of high ($\sim$10--50$M_\odot$) masses with low or zero spins \citep{The-LIGO-Scientific-Collaboration:2019} consistent with the above proposals, although alternatives have been suggested \cite{Belczynski:2016}. At present this intriguing possibility remains to be firmly ruled in - or out.  

{\it In the next decade, new space- and ground-borne  electromagnetic (EM) instruments, combined with ongoing theoretical efforts, should shed critical light on PBHs. The new data could have groundbreaking implications for DM and our understanding of the physics in the early Universe.}

{\bf Source-subtracted cosmic infrared background.}
\begin{figure}[b!]
    \centering
 \includegraphics[width=5.in]{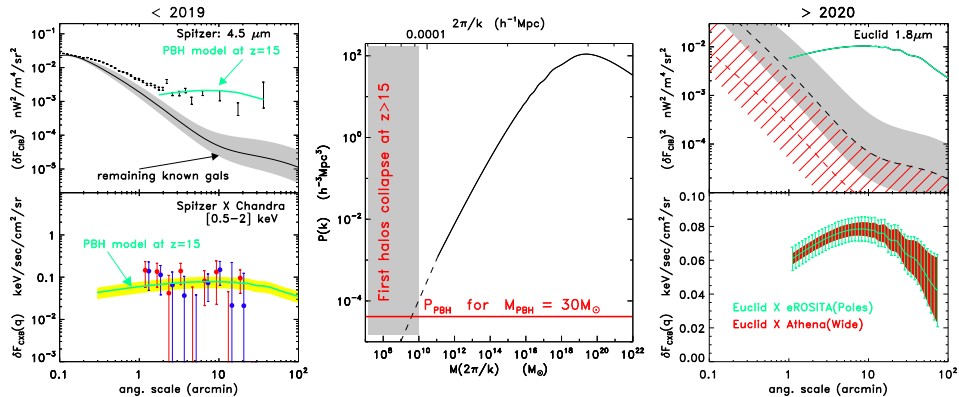}
\caption{\scriptsize {\bf Left}: Top: source-subtracted CIB anisotropies vs remaining known galaxies \cite{Kashlinsky:2012}. Bottom: reconstructed CXB fluctuation from the CIB sources are consistent with the high-$z$ origin at levels not directly detectable with future planned X-ray missions \citep{Kashlinsky:2019}. {\bf Middle}: $\Lambda$CDM power spectrum compared with the Poissonian granulation component from 30$M_\odot$ PBHs making up the DM. \citep[from ][]{Kashlinsky:2016}. {\bf Right}: Top: source-subtracted CIB fluctuations at 1.8 \mic\ due to a PBH-like model normalized to {\it Spitzer} measurements, which will be measured at sub-percent statistical accuracy with {\it Euclid} compared to that from known galaxy populations remaining in {\it Euclid}'s Wide (black) and Deep Surveys (red)\citep[adapted from][]{Kashlinsky:2018}. Bottom: CXB signal from lower left panel recovered using CIB-CXB cross-power with {\it Euclid} and {\it eROSITA} (green) and {\it Athena} (red) X-ray missions \citep{Kashlinsky:2019}.
}
\label{fig:cib}
\end{figure}
The spatial spectrum of CDM-type matter perturbations uniquely specifies the epochs and abundances of first stars era (FSE) objects at $z\gtrsim$15--20. FSE sources produce emissions in the near-IR observer bands $\lambda\!\!>$1\mic\  with potentially measurable anisotropies in the near-IR cosmic infrared background (CIB) \citep{Kashlinsky:2004,Cooray:2004}. 
Ref. \cite{Kashlinsky:2005a} have conducted the first search of this component using, with specifically developed map-making tools \citep{Arendt:2010}, deep {\it Spitzer} data at 3.6--8\mic, discovering source-subtracted CIB fluctuations significantly exceeding those from remaining known galaxies \citep{Kashlinsky:2005a,Helgason:2012a}, thereby indicating new cosmological populations. Follow-up measurements from more {\it Spitzer} fields identified the CIB fluctuation power excess to $\simeq$1$^\circ$ with similar levels across the sky \citep{Kashlinsky:2007a,Kashlinsky:2007,Kashlinsky:2012,Cooray:2012}. The shot noise of these fluctuations, $P_{\rm SN}=\int_0^{S_0}S^2dN$, is still dominated by the  known sources below the limiting flux $S_0$, while the clustering component, by the new sources, appears consistent with the high-$z$ $\Lambda$CDM  model  (Fig. \ref{fig:cib},top-left). The clustering CIB component does not yet appear to decrease with lower shot noise implying its origin in faint, possibly high-$z$, sources of $\lesssim$10--20 nJy \citep{Kashlinsky:2007b}. The source-subtracted CIB fluctuations at 3.6 and 4.5 \mic\ appear strongly coherent with soft (0.5-2 keV) unresolved cosmic X-ray background (CXB) \citep{Cappelluti:2013,Mitchell-Wynne:2016,Cappelluti:2017,Li:2018}, which cannot be accounted for by the 
remaining known populations \citep{Helgason:2014}. The CIB signal thus must arise in populations having a much higher fraction of BHs than known sources. (See review \cite{Kashlinsky:2018}.)
The standard $\Lambda$CDM power spectrum in the regime corresponding to collapsing first halos ($\sim10^{6-8}M_\odot$) at $z\gtrsim$10--15, follows $P\!\!\propto \!k^{-3}$ limiting their abundance and the resultant CIB. Yet the measured levels of the clustering component of the CIB anisotropies appear higher than simple high-$z$ evolutionary models, in conjunction with the above regime, would predict \citep{Helgason:2016} - i.e. {\it the CIB clustering component may require more power than $P\!\!\propto\!k^{-3}$}.  The excess power is naturally provided by LIGO-type PBHs and was proposed as evidence of this population \citep{Kashlinsky:2016}.
If DM is made up of LIGO-type PBHs, the latter would thus introduce an additional granulation component \citep{Meszaros:1974} of the amplitude naturally accounting for the observed source-subtracted CIB \citep{Kashlinsky:2016}, which would be Poissonian on scales of the first collapsed minihalos ($>\!\!10^{6-8}M_\odot$) \citep{Ali-Haimoud_18}. The uncovered coherence of the source-subtracted CIB with the (observer frame) soft CXB, a sign of abundant accreting BHs, would then be explained naturally by this extra component. 
Fig. \ref{fig:cib} summarizes the current measurements (left), shows the effects on the signals from LIGO-type PBHs making up DM (middle) and illustrates the future prospects for significantly more refined probes of the signal with {\it Euclid} and {\it WFIRST} CIB probes combined with upcoming X-ray missions, {\it eROSITA} and {\it Athena} (right).

{\bf Microlensing of the Galactic disk and halo}. Gravitational microlensing 
can probe the mass distribution of stellar remnants and DM in the form of compact objects, the so-called MACHOs \citep{Paczynski:1986, Griest:1991}. The technique relies on continuous monitoring of millions of stars to search for magnification events caused by gravitational lensing by massive objects crossing near the line of sight.
The observations of the Magellanic Clouds over 13 years by the MACHO, EROS and OGLE surveys as well as from other surveys, have claimed that sub-solar-mass MACHOs cannot make more than a few percent of the DM in the Milky Way halo \citep{Alcock:2001, Tisserand:2007,Wyrzykowski:2009,Wyrzykowski:2011a,Griest:2013, Calcino:2018}. 
However, in the LIGO BH mass regime there was still not enough sensitivity to confirm or rule out massive DM lenses.  
Stellar and Primordial BH lenses can be detected only in a very long and wide photometric sky surveys, e.g., MACHO \citep{Bennett:2002}, OGLE \citep{Wyrzykowski:2016}, MOA \citep{Bond:2001}, {\it Gaia} \citep{Gaia-Collaboration:2016}, VVV \citep{Navarro2018}, ASAS-SN, ZTF, etc.
Natural directions for the search are where the source stars are dense, i.e., the Galactic plane, the Magellanic Clouds or M31.
While the Galactic bulge poses a real challenge for the LSST due to crowding, the Galactic Disk (GD) fields are much sparser, yet still provide billions of potential sources for lensing.
Over its nominal operation LSST should discover hundreds of temporal lensing events due to BHs \citep{SajadianPoleski:2018}. 
Understanding the nature of the lens, however, would still require additional resources. 
High-resolution spectroscopy of the magnified source (to get the source distance) will still be a challenge at 20-23 mag range. 
\begin{figure}[b!]
\parbox{0.73\textwidth}{\hspace*{-2mm}
\includegraphics[width=0.355\textwidth,height=1.485in]{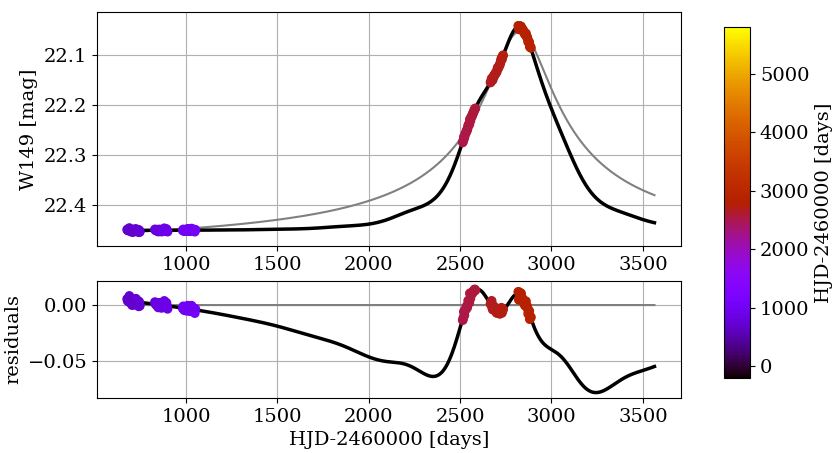}
\includegraphics[width=0.355\textwidth,height=1.485in]{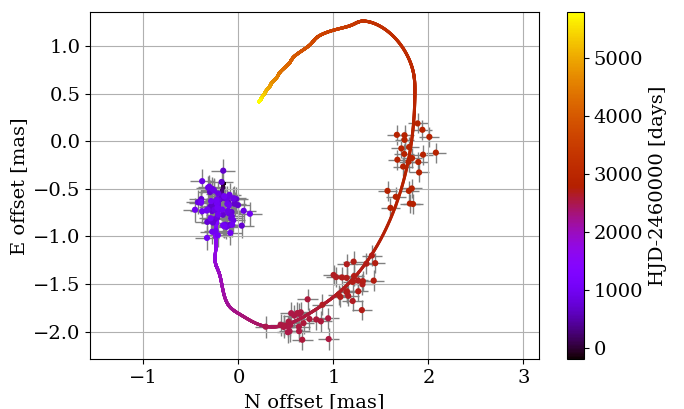}}
\hfil
\parbox{0.26\textwidth}{\vspace{-3mm}\caption{\scriptsize Example of a 9 $M_\odot$ BH microlensing event as observed by WFIRST. Left: photometric observations are marked with red points and the lines denote the standard and parallax models (thin and thick, respectively). Right: astrometric data for the source motion due to microlensing \citep{Rybicki:2018}. Color codes the time in days. 
\label{fig:wfirstlens}}}
\vspace*{-2mm}
\end{figure}
Even more important are the ultra-precise astrometric observations, as the size of the positional displacement directly yields 
the Einstein ring size, which combined with the long-term photometry, yields the lens mass measurement \citep{Lu:2016}. Milli-arcsecond angular resolution is currently possible with Gaia space mission, but with limiting magnitude of about V=16 mag \citep{Rybicki:2018}, while in future WFIRST should provide precise enough astrometric time-series (see Fig. \ref{fig:wfirstlens}). Interferometric measurements in the optical are even more brightness limited \citep{Dong:2018}.
In summary, the recipe for a BH lens discovery via microlensing is a long-term monitoring of millions of stars, assisted with spectroscopy, high-angular resolution imaging and, most importantly, detection of astrometric microlensing. 

\begin{figure}[h!]
\hspace*{2mm}
\centering
\parbox{0.6\textwidth}{
\includegraphics[width=0.45\textwidth]{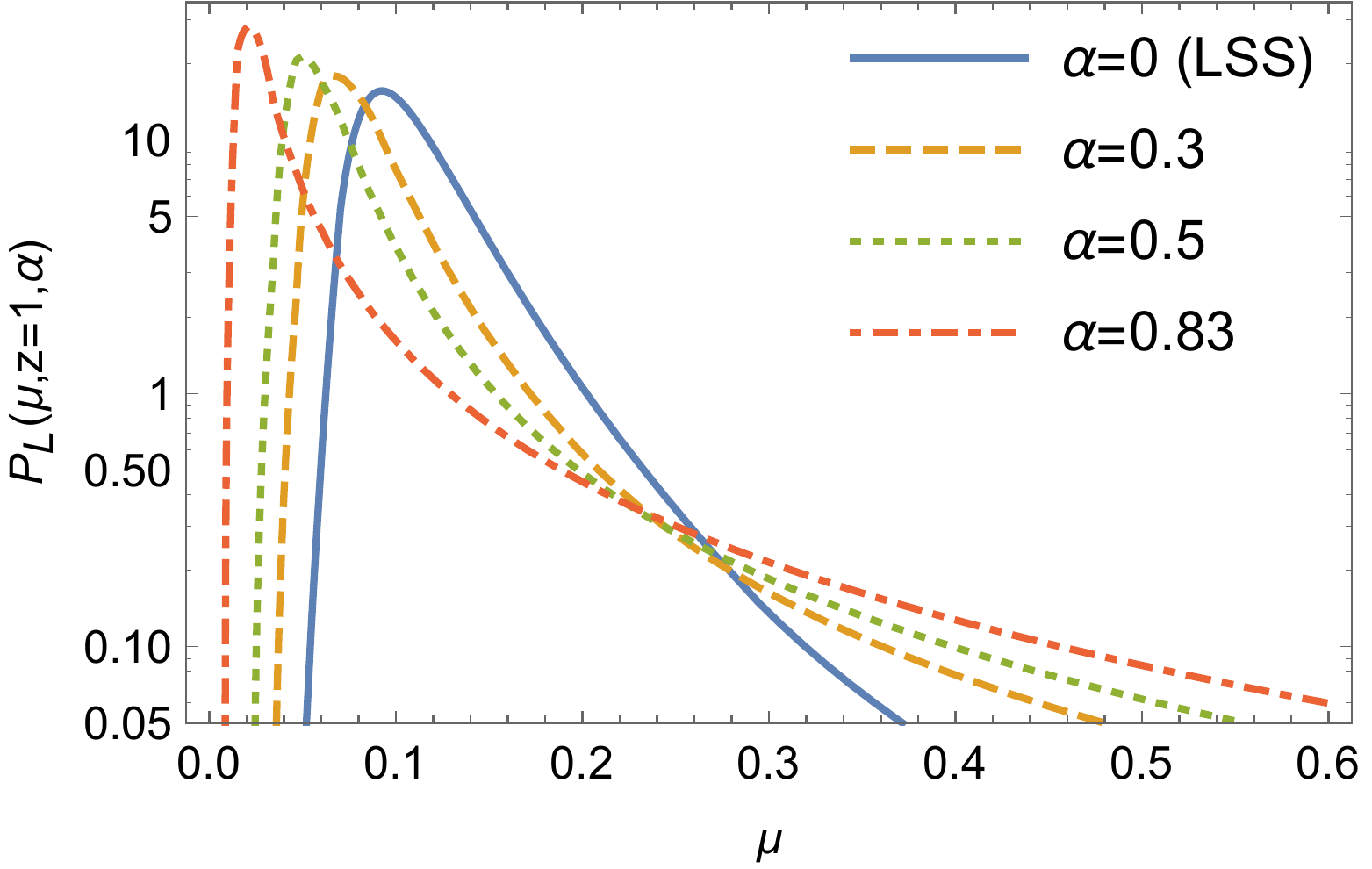}}
\hfil
\parbox{0.35\textwidth}{\vspace{-10mm}\caption{\scriptsize The probability distribution of SN magnification $\mu$ at $z$=1 for different values of the fraction $\alpha$ of DM made of compact objects ($\alpha=0.83$ corresponds to $f_{\rm PBH} = 1$ and $\alpha=0$ to $f_{\rm PBH}$=0). {\it In the future we may determine via SN lensing the abundance and concentration of DM lumps made of primordial black holes, and distinguish them from the expected substructure around galaxies in the form of dwarf spheroidals that the standard CDM hierarchical structure formation paradigm predicts.}
}
\label{fig:SNmagn}}f
\end{figure}

{\bf Supernova (SN) lensing in the upcoming surveys.}  SN lensing by intervening compact objects is a potential powerful probe of the size and mass of BHs and their clusters that could constitute the (majority of) DM \citep{Rauch:1991,Seljak:1999tm,Metcalf:1999qb,Zumalacarregui:2018,Garcia-Bellido:2018}.  Microlensing affects the distribution of SN residuals: a few events are magnified, thereby appearing brighter than the average, while the bulk is demagnified and hence appears dimmer than average (see Fig. \ref{fig:SNmagn}).  There is no evidence for such an effect in the present SN catalogs, due to their small size, and to systematic effects like host evolution (galaxy, cluster) and environment (dust, etc.)
In recent catalogs, (e.g. DES SNe \citep{Kessler:2018krb,Brout:2018}), SN light-curve fitters include models for weak lensing, but not micro-lensing. A large catalog of several thousand SNe will allow us to study all the relevant systematics and, in particular, the skewness of the residuals caused by micro-lensing. With such a catalog, one can begin to address more accurately the SN lensing signal as a function of redshift, unravelling the evolution of clustering of black holes with masses between a few and thousands of solar masses, especially if they constitute the bulk of the dark matter.

{\bf Probing PBHs with the CMB.} The Cosmic Microwave Background (CMB) is an exquisitely sensitive probe of energy injection in the early Universe. Heat injection at $10^3 \lesssim z \lesssim 2 \times 10^6$ can distort the frequency spectrum of the CMB, while energy injection at $z \lesssim 10^3$ leads to extra ionizations and increases the residual abundance of free electrons. The latter leads to a suppression of small-scale temperature and polarization power, and an enhancement of large-scale polarization fluctuations. PBHs with mass $M \gtrsim M_{\odot}$ can lead to energy injection through accretion-powered radiation~\cite{Carr:1981,Ricotti:2008}. Due to the large Compton drag by CMB photons, accretion is suppressed at early times, and the largest impact of accreting PBHs is on the recombination history \citep{AliHaimoud:2017, Aloni:2017} rather than spectral distortion \citep{Kohri:2014lza,Pani:2013hpa}. 
The most conservative  bounds are derived by assuming a modified quasi-spherical Bondi-Hoyle accretion, and a very low radiative efficiency due to free-free radiation of the infalling gas \cite{Ricotti:2008, AliHaimoud:2017}. The physics of BH accretion is rich and, at the present time, remains highly uncertain. The geometry of accretion might be disk-like rather than quasi-spherical, provided the infalling gas has sufficient angular momentum, and enough time to cool to settle into a disk. Should this be the case, the radiative efficiency might be much larger, and CMB anisotropies could be sensitive to much lighter and/or much lower abundances of PBHs \cite{Poulin:2017}. Finally, just like other probes of PBHs, accretion is sensitive to the underlying small-scale density field, as well as the clustering properties of PBHs, which remain poorly understood \cite{Chisholm_06, Ali-Haimoud_18, Ballesteros_18, Desjacques_18,Bringmann:2018mxj}. The main effort to be pursued on the topic of accreting PBHs and their impact on the CMB is therefore theoretical. In particular, the nature, rate and efficiency of early-Universe accretion onto PBHs needs to be explored in much greater depth. The problem is well defined, and significant progress can be made in the next decade by combining analytic and numerical studies. 

{\bf Other constraining EM measurements.}  
Other EM-based datasets would provide further insights into the PBH-DM connection \citep{Carr:2009jm,Carr:2016drx,Carr:2017jsz}: i) {\it Dynamical heating of star clusters in ultra-faint dwarf galaxies (UFDG)} \citep{Brandt:2016,Green:2016,Li:2016,Koushiappas:2017}. One can constrain the PBH abundance of mass above 10 $M_\odot$ within UFDG. However, the evolution and interpretation of such systems are subject to various selection effects and systematics, e.g. the assumption on the possible existence of a central massive BH, projection effects, etc. 
A way to rule out sub-solar PBH as DM is to detect an UFDG (or a stable star clusters in a UFDG) with radius $\lesssim$10 pc, which would be unstable due to dynamical heating.   Confirming a cut-off radius between 10 and 20 pc with upcoming UFDG observations could be an evidence for the compact nature of DM. ii) {\it Central DM profile in dwarf galaxies} \citep{Clesse:2018}. The scattering cross-section of solar-mass PBH is about $\sigma_{\rm PBH} / m_{\rm PBH} \approx \mathcal O (\rm cm^2/g)$, which could explain the core profiles of dwarf galaxies.  iii) {\it Measurements with pulsar timing arrays} (PTA) should probe the 2nd order Shapiro time delay from a large sample of millisecond pulsars induced by a Stochastic GW Background (SGWB) from LIGO-type BH. This background could be due to second order perturbations at PBH formation~\citep{Nakama:2016gzw,Inomata:2016rbd,Gong:2017qlj,Garcia-Bellido:2017aan,Orlofsky:2016vbd,Clesse:2018ogk}, or due to early- or late-time inspirals~\citep{Mandic:2016lcn,Clesse:2016ajp,Raidal:2017mfl}.
iv) {\it Cosmological QSO lensing variability} at various wavelengths and multi-year timescales is sensitive up to few $M_\odot$ BH~\citep{Hawkins:1993,Mediavilla:2017}. Shorter timescale microlensing of QSO/AGN and femto-lensing of GRB probe planetary mass BHs~\citep{Dai:2018,Barnacka:2012bm}.  Other probes include point sources towards the Galactic center \citep{Gaggero:2017,Hektor:2018}, wide binaries in the Galactic halo \citep{Monroy-Rodriguez:2014}, 21cm signal from reionization~\cite{Hektor:2018qqw} and strong gravitational lensing of extragalactic fast radio bursts (FRBs), which can produce observable ``echoes" of these radio signals  \citep{Munoz:2016}.\\
{\centerline{\bf Future prospects and capabilities.}}

$\bullet$ Probing the source-subtracted CIB fluctuations at high precision, determining its epochs and coherence with unresolved CXB would provide important clues to the origin of its populations and the role of PBHs in their production. The information will come from the LIBRAE ({\tiny \bf Looking at Infrared Background Radiation with {\it Euclid}}) project within {\it Euclid} \citep{site:Euclid}.  
The {\it Euclid} mission, with its deep coverage of $\sim$half the sky in three near-IR bands (Y,J,H) and visible (0.6-0.9\mic) in the Wide Survey and a still deeper coverage of additional 40deg$^2$ in the Deep Survey \cite{Laureijs:2011,Laureijs:2014}, will result, in this context, in 1) measurement of the CIB power spectrum to sub-percent statistical accuracy and establishing the epochs of its sources by cross-correlating with visible diffuse maps, and then 2) probing the CIB-CXB cross-power at high fidelity \cite{Kashlinsky:2019} with surveys from the concurrent {\it eROSITA} \cite{Merloni:2012} mission. The later  ESA's {\it Athena} X-ray mission \cite{Nandra:2013} will bring further improvements to the measurement \cite{Kashlinsky:2019} if it surveys $\sim$100deg$^2$ to the depth of the currently deepest {\it Chandra} exposures used in \cite{Cappelluti:2013,Cappelluti:2017,Mitchell-Wynne:2016,Li:2018}. See Fig.\ref{fig:cib}. Further  opportunities will be available with NASA's {\it WFIRST} \cite{Spergel:2015} mission which, while covering less area that {\it Euclid}, will go deeper and over a broader near-IR range.

$\bullet$ Microlensing: 
On-going and continuing long-term wide-field surveys like OGLE, {\it Gaia} and then primarily LSST will provide hundreds of BH lensing events. However, only with future ultra-precise and sensitive astrometric instruments, like {\it WFIRST, Euclid, JWST}, ELT or {\it Gaia}NIR, it will be possible to derive the mass of the lens, hence recognizing BH lenses.  These facilities will also detect thousands of quadruply lensed quasars, allowing to detect PBH through microlensing.

$\bullet$ SN lensing: SNeIA are the best cosmological candidates for probing PBHs via lensing. They are predicted to be $\sim$100 SNeIA/deg$^2$/yr to $z$=1 \citep{Tonry:2003}, so assembling a sufficient sample of SNIa will require a multi-year coverage of wide areas of the sky with a suitable cadence. Going to higher $z$ would increase their expected number, but would also require deeper exposures. Out to $z\!\sim$2.5 one finds the rate of $\sim$500 SNeIA/deg$^2$/yr, but with peak AB$\sim$26.5 in J,H bands \cite{Rodney:2014}. At $z\!{\gtrsim}$5 SNIa are expected to be very rare for standard formation modes \cite{Strolger:2004,Mesinger:2006,Goobar:2009}. The  {\it WFIRST} dedicated 6-month SN survey appears optimal, 
and a large catalog of several thousand SNe will be attained. The LSST Survey cadence is well-suited for detection of transients
that may turn out to be SNe
and the SNe magnification will be monitored very precisely from the expected yield of $\sim\!3\!\times\!10^5$SNIa/yr.  {\it JWST} can contribute to this effort with observations at proper cadence covering wide areas of the sky, but the current GTO programs are expected to yield several hundred SNIa to $z\!\simeq$5. On {\it Euclid} the DESIRE ({\tiny \bf Dark Energy Supernova Infra-Red Experiment}) was proposed as a dedicated 6-month NIR rolling search predicted to measure distances to 1,700 high-redshift SNe Ia out to $z\!\simeq$1.5 \cite{Astier:2014}. 

$\bullet$ The relevant CMB work in the coming decade will necessarily be mostly theoretical. Limits will moreover become more robust as more detailed understanding of the accretion process is achieved. So far CMB studies of PBHs have been limited to the CMB angular power spectrum. It is possible that higher-order statistics (such as the bispectrum) can also be used to constrain PBHs. New high-sensitivity, high-angular-resolution CMB-polarization missions,  CMB Stage-IV \cite{Abazajian:2016} and AdvACTPol \cite{The-Simons-Observatory-Collaboration:2018}, should significantly improve over \emph{Planck} for measuring cosmological parameters. A detailed study of the sensitivity of upcoming CMB-anisotropy measurements to accreting PBHs, paralleling that of \cite{Green:2018} for annihilating DM particles, needs to be completed.

$\bullet$ UFDGs:  Between hundreds and thousands of UFDG could be detected by wide and deep-field surveys with {\it Euclid}, LSST, {\it WFIRST} and {\it JWST}.   Wide surveys will have a  limiting magnitude lower than the one reached by VST-Atlas (VLT) and HCS-SSP deep field surveys.  With a limiting magnitude $r \simeq 31$, {\it JWST} will detect the faintest and smallest objects, like UFDG with half-light radius $r_{1/2}\lesssim 10$ pc and star clusters within larger UFDG that would detected by wide surveys.  

$\bullet$ Pulsar timing and FRB lensing:  The 5-year observations from the International Pulsar Timing Array (IPTA)~\cite{Verbiest:2016vem}, regrouping NANOGrav \cite{Arzoumanian:2015liz}, the Parkes Pulsar Timing Array \cite{Manchester:2012za} and the European Pulsar Timing Array \cite{Lentati:2015qwp} have a projected sensitivity to the GW density at nanohertz frequencies of $\Omega_{\rm GW} h^2 \sim 10^{-10}$~\cite{Lasky:2015lej}. With the PTA of the SKA, this sensitivity could be reduced down to  $\Omega_{\rm GW} h^2 \sim 10^{-15}$ \cite{Lazio:2013mea,Zhao:2013bba}. IPTA or SKA should detect the SGWB from (Gaussian) curvature fluctuations leading to stellar-mass PBH formation \cite{Inomata:2018epa, Byrnes:2018txb}. Upcoming FRB lensing observations from CHIME~\cite{CHIME/FRB-Collaboration:2019} can significantly constrain the DM fraction in $\gtrsim$10$M_\odot$ PBHs.

$\bullet$ In addition to the above EM investigations two inputs in this regard from the GW frontier could be critical. The golden observation that would almost surely be a sign of a PBH would be the detection of a BH merger progenitor lighter than the Chandrasekhar mass.  Searching for sub-solar mass BH is additionally motivated theoretically, because they would have formed at the QCD transition, when the sound speed reduction naturally boosts PBH formation \cite{Jedamzik:1996mr,Byrnes_18} so they may be more abundant than heavier BHs.   
In the future, it is therefore important i) to pursue the analysis of LIGO runs~\cite{Abbott:2018oah}, ii) to extend it to non-equal mass binaries (and to spinning black holes), iii) to compute more accurately the expected merger rates for different binary formation mechanisms \cite{Nakamura_97, Sasaki_16, Ali-Haimoud_17b, Raidal_18}, clustering models \cite{Bringmann_18, Desjacques_18, Ballesteros_18} and PBH mass functions, in the sub-solar mass range. Additionally in the absence of PBH contributions to DM, the standard $\Lambda$CDM matter density fluctuations precludes stars forming at very early epochs, $z\gtrsim$15--20. Finding BH mergers at progressively higher $z\gtrsim 15$, where no stars are expected in the absence of PBHs, will be a strong indicator of PBHs. The three proposed experimental configurations, Cosmic Explorer \citep{site:cosmicexplorer}, {\it LISA} \citep{site:LISA} 
and Einstein Telescope \citep{site:ET}, that can probe GWs from BH merges at such epochs, will be important for probing the PBH-DM link.

\pagebreak

\begin{thebibliography}{127}
\providecommand{\natexlab}[1]{#1}
\providecommand{\url}[1]{\texttt{#1}}
\expandafter\ifx\csname urlstyle\endcsname\relax
  \providecommand{\doi}[1]{doi: #1}\else
  \providecommand{\doi}{doi: \begingroup \urlstyle{rm}\Url}\fi

\bibitem[{Abazajian} et~al.(2016){Abazajian}, {Adshead}, {Ahmed}, {Allen},
  {Alonso}, {Arnold}, {Baccigalupi}, {Bartlett}, {Battaglia}, {Benson},
  {Bischoff}, {Borrill}, {Buza}, {Calabrese}, {Caldwell}, {Carlstrom}, {Chang},
  {Crawford}, {Cyr-Racine}, {De Bernardis}, {de Haan}, {di Serego Alighieri},
  {Dunkley}, {Dvorkin}, {Errard}, {Fabbian}, {Feeney}, {Ferraro}, {Filippini},
  {Flauger}, {Fuller}, {Gluscevic}, {Green}, {Grin}, {Grohs}, {Henning},
  {Hill}, {Hlozek}, {Holder}, {Holzapfel}, {Hu}, {Huffenberger}, {Keskitalo},
  {Knox}, {Kosowsky}, {Kovac}, {Kovetz}, {Kuo}, {Kusaka}, {Le Jeune}, {Lee},
  {Lilley}, {Loverde}, {Madhavacheril}, {Mantz}, {Marsh}, {McMahon},
  {Meerburg}, {Meyers}, {Miller}, {Munoz}, {Nguyen}, {Niemack}, {Peloso},
  {Peloton}, {Pogosian}, {Pryke}, {Raveri}, {Reichardt}, {Rocha}, {Rotti},
  {Schaan}, {Schmittfull}, {Scott}, {Sehgal}, {Shandera}, {Sherwin}, {Smith},
  {Sorbo}, {Starkman}, {Story}, {van Engelen}, {Vieira}, {Watson}, {Whitehorn},
  and {Kimmy Wu}]{Abazajian:2016}
K.~N. {Abazajian}, P.~{Adshead}, Z.~{Ahmed}, S.~W. {Allen}, D.~{Alonso}, K.~S.
  {Arnold}, C.~{Baccigalupi}, J.~G. {Bartlett}, N.~{Battaglia}, B.~A. {Benson},
  C.~A. {Bischoff}, J.~{Borrill}, V.~{Buza}, E.~{Calabrese}, R.~{Caldwell},
  J.~E. {Carlstrom}, C.~L. {Chang}, T.~M. {Crawford}, F.-Y. {Cyr-Racine},
  F.~{De Bernardis}, T.~{de Haan}, S.~{di Serego Alighieri}, J.~{Dunkley},
  C.~{Dvorkin}, J.~{Errard}, G.~{Fabbian}, S.~{Feeney}, S.~{Ferraro}, J.~P.
  {Filippini}, R.~{Flauger}, G.~M. {Fuller}, V.~{Gluscevic}, D.~{Green},
  D.~{Grin}, E.~{Grohs}, J.~W. {Henning}, J.~C. {Hill}, R.~{Hlozek},
  G.~{Holder}, W.~{Holzapfel}, W.~{Hu}, K.~M. {Huffenberger}, R.~{Keskitalo},
  L.~{Knox}, A.~{Kosowsky}, J.~{Kovac}, E.~D. {Kovetz}, C.-L. {Kuo},
  A.~{Kusaka}, M.~{Le Jeune}, A.~T. {Lee}, M.~{Lilley}, M.~{Loverde}, M.~S.
  {Madhavacheril}, A.~{Mantz}, D.~J.~E. {Marsh}, J.~{McMahon}, P.~D.
  {Meerburg}, J.~{Meyers}, A.~D. {Miller}, J.~B. {Munoz}, H.~N. {Nguyen}, M.~D.
  {Niemack}, M.~{Peloso}, J.~{Peloton}, L.~{Pogosian}, C.~{Pryke}, M.~{Raveri},
  C.~L. {Reichardt}, G.~{Rocha}, A.~{Rotti}, E.~{Schaan}, M.~M. {Schmittfull},
  D.~{Scott}, N.~{Sehgal}, S.~{Shandera}, B.~D. {Sherwin}, T.~L. {Smith},
  L.~{Sorbo}, G.~D. {Starkman}, K.~T. {Story}, A.~{van Engelen}, J.~D.
  {Vieira}, S.~{Watson}, N.~{Whitehorn}, and W.~L. {Kimmy Wu}.
\newblock {CMB-S4 Science Book, First Edition}.
\newblock \emph{ArXiv e-prints}, October 2016.

\bibitem[{Abbott} et~al.(2016{\natexlab{a}}){Abbott}, {Abbott}, {Abbott},
  {Abernathy}, {Acernese}, {Ackley}, {Adams}, {Adams}, {Addesso}, {Adhikari},
  and et~al.]{Abbott:2016b}
B.~P. {Abbott}, R.~{Abbott}, T.~D. {Abbott}, M.~R. {Abernathy}, F.~{Acernese},
  K.~{Ackley}, C.~{Adams}, T.~{Adams}, P.~{Addesso}, R.~X. {Adhikari}, and
  et~al.
\newblock {GW151226: Observation of Gravitational Waves from a 22-Solar-Mass
  Binary Black Hole Coalescence}.
\newblock \emph{Physical Review Letters}, 116\penalty0 (24):\penalty0 241103,
  June 2016{\natexlab{a}}.
\newblock \doi{10.1103/PhysRevLett.116.241103}.

\bibitem[{Abbott} et~al.(2016{\natexlab{b}}){Abbott}, {Abbott}, {Abbott},
  {Abernathy}, {Acernese}, {Ackley}, {Adams}, {Adams}, {Addesso}, {Adhikari},
  and et~al.]{Abbott:2016c}
B.~P. {Abbott}, R.~{Abbott}, T.~D. {Abbott}, M.~R. {Abernathy}, F.~{Acernese},
  K.~{Ackley}, C.~{Adams}, T.~{Adams}, P.~{Addesso}, R.~X. {Adhikari}, and
  et~al.
\newblock {GW150914: First results from the search for binary black hole
  coalescence with Advanced LIGO}.
\newblock \emph{\prd}, 93\penalty0 (12):\penalty0 122003, June
  2016{\natexlab{b}}.
\newblock \doi{10.1103/PhysRevD.93.122003}.

\bibitem[Abbott et~al.(2018)]{Abbott:2018oah}
B.~P. Abbott et~al.
\newblock {Search for Subsolar-Mass Ultracompact Binaries in Advanced LIGO?s
  First Observing Run}.
\newblock \emph{Phys. Rev. Lett.}, 121\penalty0 (23):\penalty0 231103, 2018.
\newblock \doi{10.1103/PhysRevLett.121.231103}.

\bibitem[{Alcock} et~al.(2001){Alcock}, {Allsman}, {Alves}, {Axelrod},
  {Becker}, {Bennett}, {Cook}, {Dalal}, {Drake}, {Freeman}, {Geha}, {Griest},
  {Lehner}, {Marshall}, {Minniti}, {Nelson}, {Peterson}, {Popowski}, {Pratt},
  {Quinn}, {Stubbs}, {Sutherland}, {Tomaney}, {Vandehei}, and
  {Welch}]{Alcock:2001}
C.~{Alcock}, R.~A. {Allsman}, D.~R. {Alves}, T.~S. {Axelrod}, A.~C. {Becker},
  D.~P. {Bennett}, K.~H. {Cook}, N.~{Dalal}, A.~J. {Drake}, K.~C. {Freeman},
  M.~{Geha}, K.~{Griest}, M.~J. {Lehner}, S.~L. {Marshall}, D.~{Minniti}, C.~A.
  {Nelson}, B.~A. {Peterson}, P.~{Popowski}, M.~R. {Pratt}, P.~J. {Quinn},
  C.~W. {Stubbs}, W.~{Sutherland}, A.~B. {Tomaney}, T.~{Vandehei}, and D.~L.
  {Welch}.
\newblock {MACHO Project Limits on Black Hole Dark Matter in the 1-30
  M$_{solar}$ Range}.
\newblock \emph{\apj}, 550:\penalty0 L169--L172, April 2001.
\newblock \doi{10.1086/319636}.

\bibitem[{Ali-Ha{\"\i}moud}(2018)]{Ali-Haimoud_18}
Y.~{Ali-Ha{\"\i}moud}.
\newblock {Correlation Function of High-Threshold Regions and Application to
  the Initial Small-Scale Clustering of Primordial Black Holes}.
\newblock \emph{Physical Review Letters}, 121\penalty0 (8):\penalty0 081304,
  August 2018.
\newblock \doi{10.1103/PhysRevLett.121.081304}.

\bibitem[{Ali-Ha{\"\i}moud} and {Kamionkowski}(2017)]{AliHaimoud:2017}
Y.~{Ali-Ha{\"\i}moud} and M.~{Kamionkowski}.
\newblock {Cosmic microwave background limits on accreting primordial black
  holes}.
\newblock \emph{\prd}, 95\penalty0 (4):\penalty0 043534, February 2017.
\newblock \doi{10.1103/PhysRevD.95.043534}.

\bibitem[{Ali-Ha{\"i}moud} et~al.(2017){Ali-Ha{\"i}moud}, {Kovetz}, and
  {Kamionkowski}]{Ali-Haimoud_17b}
Y.~{Ali-Ha{\"i}moud}, E.~D. {Kovetz}, and M.~{Kamionkowski}.
\newblock {Merger rate of primordial black-hole binaries}.
\newblock \emph{\prd}, 96\penalty0 (12):\penalty0 123523, December 2017.
\newblock \doi{10.1103/PhysRevD.96.123523}.

\bibitem[{Aloni} et~al.(2017){Aloni}, {Blum}, and {Flauger}]{Aloni:2017}
D.~{Aloni}, K.~{Blum}, and R.~{Flauger}.
\newblock {Cosmic microwave background constraints on primordial black hole
  dark matter}.
\newblock \emph{JCAP}, 5:\penalty0 017, May 2017.
\newblock \doi{10.1088/1475-7516/2017/05/017}.

\bibitem[Antenna()]{site:LISA}
Laser Interferometer~Space Antenna.
\newblock URL \url{https://lisa.nasa.gov/}.

\bibitem[{Arendt} et~al.(2010){Arendt}, {Kashlinsky}, {Moseley}, and
  {Mather}]{Arendt:2010}
R.~G. {Arendt}, A.~{Kashlinsky}, S.~H. {Moseley}, and J.~{Mather}.
\newblock {Cosmic Infrared Background Fluctuations in Deep Spitzer Infrared
  Array Camera Images: Data Processing and Analysis}.
\newblock \emph{\apj}, 186:\penalty0 10--47, January 2010.
\newblock \doi{10.1088/0067-0049/186/1/10}.

\bibitem[Arzoumanian et~al.(2016)]{Arzoumanian:2015liz}
Z.~Arzoumanian et~al.
\newblock {The NANOGrav Nine-year Data Set: Limits on the Isotropic Stochastic
  Gravitational Wave Background}.
\newblock \emph{Astrophys. J.}, 821\penalty0 (1):\penalty0 13, 2016.
\newblock \doi{10.3847/0004-637X/821/1/13}.

\bibitem[{Astier} et~al.(2014){Astier}, {Balland}, {Brescia}, {Cappellaro},
  {Carlberg}, {Cavuoti}, {Della Valle}, {Gangler}, {Goobar}, {Guy}, {Hardin},
  {Hook}, {Kessler}, {Kim}, {Linder}, {Longo}, {Maguire}, {Mannucci},
  {Mattila}, {Nichol}, {Pain}, {Regnault}, {Spiro}, {Sullivan}, {Tao},
  {Turatto}, {Wang}, and {Wood-Vasey}]{Astier:2014}
P.~{Astier}, C.~{Balland}, M.~{Brescia}, E.~{Cappellaro}, R.~G. {Carlberg},
  S.~{Cavuoti}, M.~{Della Valle}, E.~{Gangler}, A.~{Goobar}, J.~{Guy},
  D.~{Hardin}, I.~M. {Hook}, R.~{Kessler}, A.~{Kim}, E.~{Linder}, G.~{Longo},
  K.~{Maguire}, F.~{Mannucci}, S.~{Mattila}, R.~{Nichol}, R.~{Pain},
  N.~{Regnault}, S.~{Spiro}, M.~{Sullivan}, C.~{Tao}, M.~{Turatto}, X.~F.
  {Wang}, and W.~M. {Wood-Vasey}.
\newblock {Extending the supernova Hubble diagram to z \~{} 1.5 with the Euclid
  space mission}.
\newblock \emph{\aap}, 572:\penalty0 A80, December 2014.
\newblock \doi{10.1051/0004-6361/201423551}.

\bibitem[{Ballesteros} et~al.(2018){Ballesteros}, {Serpico}, and
  {Taoso}]{Ballesteros_18}
G.~{Ballesteros}, P.~D. {Serpico}, and M.~{Taoso}.
\newblock {On the merger rate of primordial black holes: effects of nearest
  neighbours distribution and clustering}.
\newblock \emph{JCAP}, 10:\penalty0 043, October 2018.
\newblock \doi{10.1088/1475-7516/2018/10/043}.

\bibitem[Barnacka et~al.(2012)Barnacka, Glicenstein, and
  Moderski]{Barnacka:2012bm}
A.~Barnacka, J.~F. Glicenstein, and R.~Moderski.
\newblock {New constraints on primordial black holes abundance from
  femtolensing of gamma-ray bursts}.
\newblock \emph{Phys. Rev.}, D86:\penalty0 043001, 2012.
\newblock \doi{10.1103/PhysRevD.86.043001}.

\bibitem[{Belczynski} et~al.(2016){Belczynski}, {Holz}, {Bulik}, and
  {O'Shaughnessy}]{Belczynski:2016}
K.~{Belczynski}, D.~E. {Holz}, T.~{Bulik}, and R.~{O'Shaughnessy}.
\newblock {The first gravitational-wave source from the isolated evolution of
  two stars in the 40-100 solar mass range}.
\newblock \emph{\nat}, 534:\penalty0 512--515, June 2016.
\newblock \doi{10.1038/nature18322}.

\bibitem[{Bennett} et~al.(2002){Bennett}, {Becker}, {Quinn}, {Tomaney},
  {Alcock}, {Allsman}, {Alves}, {Axelrod}, {Calitz}, {Cook}, {Drake},
  {Fragile}, {Freeman}, {Geha}, {Griest}, {Johnson}, {Keller}, {Laws},
  {Lehner}, {Marshall}, {Minniti}, {Nelson}, {Peterson}, {Popowski}, {Pratt},
  {Quinn}, {Rhie}, {Stubbs}, {Sutherland}, {Vandehei}, {Welch}, {MACHO
  Collaboration}, and {MPS Collaboration}]{Bennett:2002}
D.~P. {Bennett}, A.~C. {Becker}, J.~L. {Quinn}, A.~B. {Tomaney}, C.~{Alcock},
  R.~A. {Allsman}, D.~R. {Alves}, T.~S. {Axelrod}, J.~J. {Calitz}, K.~H.
  {Cook}, A.~J. {Drake}, P.~C. {Fragile}, K.~C. {Freeman}, M.~{Geha},
  K.~{Griest}, B.~R. {Johnson}, S.~C. {Keller}, C.~{Laws}, M.~J. {Lehner},
  S.~L. {Marshall}, D.~{Minniti}, C.~A. {Nelson}, B.~A. {Peterson},
  P.~{Popowski}, M.~R. {Pratt}, P.~J. {Quinn}, S.~H. {Rhie}, C.~W. {Stubbs},
  W.~{Sutherland}, T.~{Vandehei}, D.~{Welch}, {MACHO Collaboration}, and {MPS
  Collaboration}.
\newblock {Gravitational Microlensing Events Due to Stellar-Mass Black Holes}.
\newblock \emph{\apj}, 579:\penalty0 639--659, November 2002.
\newblock \doi{10.1086/342225}.

\bibitem[{Bird} et~al.(2016){Bird}, {Cholis}, {Mu{\~n}oz}, {Ali-Ha{\"\i}moud},
  {Kamionkowski}, {Kovetz}, {Raccanelli}, and {Riess}]{Bird:2016}
S.~{Bird}, I.~{Cholis}, J.~B. {Mu{\~n}oz}, Y.~{Ali-Ha{\"\i}moud},
  M.~{Kamionkowski}, E.~D. {Kovetz}, A.~{Raccanelli}, and A.~G. {Riess}.
\newblock {Did LIGO Detect Dark Matter?}
\newblock \emph{Physical Review Letters}, 116\penalty0 (20):\penalty0 201301,
  May 2016.
\newblock \doi{10.1103/PhysRevLett.116.201301}.

\bibitem[{Bond} et~al.(2001){Bond}, {Abe}, {Dodd}, {Hearnshaw}, {Honda},
  {Jugaku}, {Kilmartin}, {Marles}, {Masuda}, {Matsubara}, {Muraki}, {Nakamura},
  {Nankivell}, {Noda}, {Noguchi}, {Ohnishi}, {Rattenbury}, {Reid}, {Saito},
  {Sato}, {Sekiguchi}, {Skuljan}, {Sullivan}, {Sumi}, {Takeuti}, {Watase},
  {Wilkinson}, {Yamada}, {Yanagisawa}, and {Yock}]{Bond:2001}
I.~A. {Bond}, F.~{Abe}, R.~J. {Dodd}, J.~B. {Hearnshaw}, M.~{Honda},
  J.~{Jugaku}, P.~M. {Kilmartin}, A.~{Marles}, K.~{Masuda}, Y.~{Matsubara},
  Y.~{Muraki}, T.~{Nakamura}, G.~{Nankivell}, S.~{Noda}, C.~{Noguchi},
  K.~{Ohnishi}, N.~J. {Rattenbury}, M.~{Reid}, T.~{Saito}, H.~{Sato},
  M.~{Sekiguchi}, J.~{Skuljan}, D.~J. {Sullivan}, T.~{Sumi}, M.~{Takeuti},
  Y.~{Watase}, S.~{Wilkinson}, R.~{Yamada}, T.~{Yanagisawa}, and P.~C.~M.
  {Yock}.
\newblock {Real-time difference imaging analysis of MOA Galactic bulge
  observations during 2000}.
\newblock \emph{\mnras}, 327:\penalty0 868--880, November 2001.
\newblock \doi{10.1046/j.1365-8711.2001.04776.x}.

\bibitem[Brandt(2016)]{Brandt:2016}
Timothy~D. Brandt.
\newblock {Constraints on MACHO Dark Matter from Compact Stellar Systems in
  Ultra-Faint Dwarf Galaxies}.
\newblock \emph{Astrophys. J.}, 824\penalty0 (2):\penalty0 L31, 2016.
\newblock \doi{10.3847/2041-8205/824/2/L31}.

\bibitem[Bringmann et~al.(2018{\natexlab{a}})Bringmann, Depta, Domcke, and
  Schmidt-Hoberg]{Bringmann:2018mxj}
Torsten Bringmann, Paul~Frederik Depta, Valerie Domcke, and Kai Schmidt-Hoberg.
\newblock {Strong constraints on clustered primordial black holes as dark
  matter}.
\newblock 2018{\natexlab{a}}.

\bibitem[Bringmann et~al.(2018{\natexlab{b}})Bringmann, Depta, Domcke, and
  Schmidt-Hoberg]{Bringmann_18}
Torsten Bringmann, Paul~Frederik Depta, Valerie Domcke, and Kai Schmidt-Hoberg.
\newblock {Strong constraints on clustered primordial black holes as dark
  matter}.
\newblock 2018{\natexlab{b}}.

\bibitem[Brout et~al.(2018)]{Brout:2018}
D.~Brout et~al.
\newblock {First Cosmology Results Using Type Ia Supernovae From the Dark
  Energy Survey: Photometric Pipeline and Light Curve Data Release}.
\newblock \emph{arXiv:1811.02378}, 2018.

\bibitem[Byrnes et~al.(2018{\natexlab{a}})Byrnes, Cole, and
  Patil]{Byrnes:2018txb}
Christian~T. Byrnes, Philippa~S. Cole, and Subodh~P. Patil.
\newblock {Steepest growth of the power spectrum and primordial black holes}.
\newblock 2018{\natexlab{a}}.

\bibitem[Byrnes et~al.(2018{\natexlab{b}})Byrnes, Hindmarsh, Young, and
  Hawkins]{Byrnes_18}
Christian~T. Byrnes, Mark Hindmarsh, Sam Young, and Michael~R.S. Hawkins.
\newblock Primordial black holes with an accurate {QCD} equation of state.
\newblock \emph{Journal of Cosmology and Astroparticle Physics}, 2018\penalty0
  (08):\penalty0 041--041, aug 2018{\natexlab{b}}.
\newblock \doi{10.1088/1475-7516/2018/08/041}.

\bibitem[Calcino et~al.(2018)Calcino, Garc\'ia-Bellido, and
  Davis]{Calcino:2018}
Josh Calcino, Juan Garc\'ia-Bellido, and Tamara~M. Davis.
\newblock {Updating the MACHO fraction of the Milky Way dark halo with improved
  mass models}.
\newblock \emph{Mon. Not. Roy. Astron. Soc.}, 479\penalty0 (3):\penalty0
  2889--2905, 2018.
\newblock \doi{10.1093/mnras/sty1368}.

\bibitem[{Cappelluti} et~al.(2013){Cappelluti}, {Kashlinsky}, {Arendt},
  {Comastri}, {Fazio}, {Finoguenov}, {Hasinger}, {Mather}, {Miyaji}, and
  {Moseley}]{Cappelluti:2013}
N.~{Cappelluti}, A.~{Kashlinsky}, R.~G. {Arendt}, A.~{Comastri}, G.~G. {Fazio},
  A.~{Finoguenov}, G.~{Hasinger}, J.~C. {Mather}, T.~{Miyaji}, and S.~H.
  {Moseley}.
\newblock {Cross-correlating Cosmic Infrared and X-Ray Background Fluctuations:
  Evidence of Significant Black Hole Populations among the CIB Sources}.
\newblock \emph{\apj}, 769:\penalty0 68, May 2013.
\newblock \doi{10.1088/0004-637X/769/1/68}.

\bibitem[{Cappelluti} et~al.(2017){Cappelluti}, {Arendt}, {Kashlinsky}, {Li},
  {Hasinger}, {Helgason}, {Urry}, {Natarajan}, and
  {Finoguenov}]{Cappelluti:2017}
N.~{Cappelluti}, R.~{Arendt}, A.~{Kashlinsky}, Y.~{Li}, G.~{Hasinger},
  K.~{Helgason}, M.~{Urry}, P.~{Natarajan}, and A.~{Finoguenov}.
\newblock {Probing Large-scale Coherence between Spitzer IR and Chandra X-Ray
  Source-subtracted Cosmic Backgrounds}.
\newblock \emph{\apj}, 847:\penalty0 L11, September 2017.
\newblock \doi{10.3847/2041-8213/aa8acd}.

\bibitem[{Carr}(1981)]{Carr:1981}
B.~J. {Carr}.
\newblock {Pregalactic black hole accretion and the thermal history of the
  universe}.
\newblock \emph{MNRAS}, 194:\penalty0 639--668, February 1981.
\newblock \doi{10.1093/mnras/194.3.639}.

\bibitem[Carr et~al.(2010)Carr, Kohri, Sendouda, and Yokoyama]{Carr:2009jm}
B.~J. Carr, Kazunori Kohri, Yuuiti Sendouda, and Jun'ichi Yokoyama.
\newblock {New cosmological constraints on primordial black holes}.
\newblock \emph{Phys. Rev.}, D81:\penalty0 104019, 2010.
\newblock \doi{10.1103/PhysRevD.81.104019}.

\bibitem[Carr et~al.(2016)Carr, Kuhnel, and Sandstad]{Carr:2016drx}
Bernard Carr, Florian Kuhnel, and Marit Sandstad.
\newblock {Primordial Black Holes as Dark Matter}.
\newblock \emph{Phys. Rev.}, D94\penalty0 (8):\penalty0 083504, 2016.
\newblock \doi{10.1103/PhysRevD.94.083504}.

\bibitem[Carr et~al.(2017)Carr, Raidal, Tenkanen, Vaskonen, and
  Veermae]{Carr:2017jsz}
Bernard Carr, Martti Raidal, Tommi Tenkanen, Ville Vaskonen, and Hardi Veermae.
\newblock {Primordial black hole constraints for extended mass functions}.
\newblock \emph{Phys. Rev.}, D96\penalty0 (2):\penalty0 023514, 2017.
\newblock \doi{10.1103/PhysRevD.96.023514}.

\bibitem[Carr and Hawking(1974)]{Carr:1974nx}
Bernard~J. Carr and S.~W. Hawking.
\newblock {Black holes in the early Universe}.
\newblock \emph{Mon. Not. Roy. Astron. Soc.}, 168:\penalty0 399--415, 1974.

\bibitem[{Chapline}(1975)]{1975Natur.253..251C}
G.~F. {Chapline}.
\newblock {Cosmological effects of primordial black holes}.
\newblock \emph{\nat}, 253:\penalty0 251, January 1975.
\newblock \doi{10.1038/253251a0}.

\bibitem[{CHIME/FRB Collaboration} et~al.(2019){CHIME/FRB Collaboration}, {:},
  {Amiri}, {Bandura}, {Bhardwaj}, {Boubel}, {Boyce}, {Boyle}, {Brar},
  {Burhanpurkar}, {Chawla}, {Cliche}, {Cubranic}, {Deng}, {Denman}, {Dobbs},
  {Fandino}, {Fonseca}, {Gaensler}, {Gilbert}, {Giri}, {Good}, {Halpern},
  {Hanna}, {Hill}, {Hinshaw}, {H{\"o}fer}, {Josephy}, {Kaspi}, {Landecker},
  {Lang}, {Masui}, {Mckinven}, {Mena-Parra}, {Merryfield}, {Milutinovic},
  {Moatti}, {Naidu}, {Newburgh}, {Ng}, {Patel}, {Pen}, {Pinsonneault-Marotte},
  {Pleunis}, {Rafiei-Ravandi}, {Ransom}, {Renard}, {Scholz}, {Shaw}, {Siegel},
  {Smith}, {Stairs}, {Tendulkar}, {Tretyakov}, {Vanderlinde}, and
  {Yadav}]{CHIME/FRB-Collaboration:2019}
{CHIME/FRB Collaboration}, {:}, M.~{Amiri}, K.~{Bandura}, M.~{Bhardwaj},
  P.~{Boubel}, M.~M. {Boyce}, P.~J. {Boyle}, C.~{Brar}, M.~{Burhanpurkar},
  P.~{Chawla}, J.~F. {Cliche}, D.~{Cubranic}, M.~{Deng}, N.~{Denman},
  M.~{Dobbs}, M.~{Fandino}, E.~{Fonseca}, B.~M. {Gaensler}, A.~J. {Gilbert},
  U.~{Giri}, D.~C. {Good}, M.~{Halpern}, D.~{Hanna}, A.~S. {Hill},
  G.~{Hinshaw}, C.~{H{\"o}fer}, A.~{Josephy}, V.~M. {Kaspi}, T.~L. {Landecker},
  D.~A. {Lang}, K.~W. {Masui}, R.~{Mckinven}, J.~{Mena-Parra}, M.~{Merryfield},
  N.~{Milutinovic}, C.~{Moatti}, A.~{Naidu}, L.~B. {Newburgh}, C.~{Ng},
  C.~{Patel}, U.-L. {Pen}, T.~{Pinsonneault-Marotte}, Z.~{Pleunis},
  M.~{Rafiei-Ravandi}, S.~M. {Ransom}, A.~{Renard}, P.~{Scholz}, J.~R. {Shaw},
  S.~R. {Siegel}, K.~M. {Smith}, I.~H. {Stairs}, S.~P. {Tendulkar},
  I.~{Tretyakov}, K.~{Vanderlinde}, and P.~{Yadav}.
\newblock {Observations of Fast Radio Bursts at Frequencies down to 400
  Megahertz}.
\newblock \emph{arXiv e-prints}, January 2019.

\bibitem[{Chisholm}(2006)]{Chisholm_06}
J.~R. {Chisholm}.
\newblock {Clustering of primordial black holes: Basic results}.
\newblock \emph{\prd}, 73\penalty0 (8):\penalty0 083504, April 2006.
\newblock \doi{10.1103/PhysRevD.73.083504}.

\bibitem[{Clesse} and {Garc{\'{\i}}a-Bellido}(2017)]{Clesse:2017}
S.~{Clesse} and J.~{Garc{\'{\i}}a-Bellido}.
\newblock {The clustering of massive Primordial Black Holes as Dark Matter:
  Measuring their mass distribution with advanced LIGO}.
\newblock \emph{Physics of the Dark Universe}, 15:\penalty0 142--147, 2017.
\newblock \doi{10.1016/j.dark.2016.10.002}.

\bibitem[{Clesse} and {Garc{\'{\i}}a-Bellido}(2018)]{Clesse:2018}
S.~{Clesse} and J.~{Garc{\'{\i}}a-Bellido}.
\newblock {Seven hints for primordial black hole dark matter}.
\newblock \emph{Physics of the Dark Universe}, 22:\penalty0 137--146, December
  2018.
\newblock \doi{10.1016/j.dark.2018.08.004}.

\bibitem[Clesse and Garc{\'\i}a-Bellido(2017)]{Clesse:2016ajp}
Sebastien Clesse and Juan Garc{\'\i}a-Bellido.
\newblock {Detecting the gravitational wave background from primordial black
  hole dark matter}.
\newblock \emph{Phys. Dark Univ.}, 18:\penalty0 105--114, 2017.
\newblock \doi{10.1016/j.dark.2017.10.001}.

\bibitem[Clesse et~al.(2018)Clesse, Garc{\'\i}a-Bellido, and
  Orani]{Clesse:2018ogk}
Sebastien Clesse, Juan Garc{\'\i}a-Bellido, and Stefano Orani.
\newblock {Detecting the Stochastic Gravitational Wave Background from
  Primordial Black Hole Formation}.
\newblock \emph{arXiv:1812.11011}, 2018.

\bibitem[{Cooray} et~al.(2004){Cooray}, {Bock}, {Keatin}, {Lange}, and
  {Matsumoto}]{Cooray:2004}
A.~{Cooray}, J.~J. {Bock}, B.~{Keatin}, A.~E. {Lange}, and T.~{Matsumoto}.
\newblock {First Star Signature in Infrared Background Anisotropies}.
\newblock \emph{\apj}, 606:\penalty0 611--624, May 2004.
\newblock \doi{10.1086/383137}.

\bibitem[{Cooray} et~al.(2012){Cooray}, {Smidt}, {de Bernardis}, {Gong},
  {Stern}, {Ashby}, {Eisenhardt}, {Frazer}, {Gonzalez}, {Kochanek},
  {Koz{\l}owski}, and {Wright}]{Cooray:2012}
A.~{Cooray}, J.~{Smidt}, F.~{de Bernardis}, Y.~{Gong}, D.~{Stern}, M.~L.~N.
  {Ashby}, P.~R. {Eisenhardt}, C.~C. {Frazer}, A.~H. {Gonzalez}, C.~S.
  {Kochanek}, S.~{Koz{\l}owski}, and E.~L. {Wright}.
\newblock {Near-infrared background anisotropies from diffuse intrahalo light
  of galaxies}.
\newblock \emph{\nat}, 490:\penalty0 514--516, October 2012.
\newblock \doi{10.1038/nature11474}.

\bibitem[{Dai} and {Guerras}(2018)]{Dai:2018}
X.~{Dai} and E.~{Guerras}.
\newblock {Probing Extragalactic Planets Using Quasar Microlensing}.
\newblock \emph{\apj}, 853:\penalty0 L27, February 2018.
\newblock \doi{10.3847/2041-8213/aaa5fb}.

\bibitem[{Desjacques} and {Riotto}(2018)]{Desjacques_18}
V.~{Desjacques} and A.~{Riotto}.
\newblock {Spatial clustering of primordial black holes}.
\newblock \emph{\prd}, 98\penalty0 (12):\penalty0 123533, December 2018.
\newblock \doi{10.1103/PhysRevD.98.123533}.

\bibitem[{Dong} et~al.(2018){Dong}, {M{\'e}rand}, {Delplancke-Str{\"o}bele},
  {Gould}, {Chen}, {Post}, {Kochanek}, {Stanek}, {Christie}, {Mutel},
  {Natusch}, {Holoien}, {Prieto}, {Shappee}, and {Thompson}]{Dong:2018}
S.~{Dong}, A.~{M{\'e}rand}, F.~{Delplancke-Str{\"o}bele}, A.~{Gould},
  P.~{Chen}, R.~{Post}, C.~S. {Kochanek}, K.~Z. {Stanek}, G.~W. {Christie},
  R.~{Mutel}, T.~{Natusch}, T.~W.-S. {Holoien}, J.~L. {Prieto}, B.~J.
  {Shappee}, and T.~A. {Thompson}.
\newblock {First Resolution of Microlensed Images}.
\newblock \emph{ArXiv e-prints}, September 2018.

\bibitem[Euclid()]{site:Euclid}
Euclid.
\newblock URL \url{https://www.euclid.caltech.edu/page/Kashlinsky%20Team}.

\bibitem[Explorer()]{site:cosmicexplorer}
Cosmic Explorer.
\newblock URL \url{http://www.cosmicexplorer.org/}.

\bibitem[Gaggero et~al.(2017)Gaggero, Bertone, Calore, Connors, Lovell,
  Markoff, and Storm]{Gaggero:2017}
Daniele Gaggero, Gianfranco Bertone, Francesca Calore, Riley M.~T. Connors,
  Mark Lovell, Sera Markoff, and Emma Storm.
\newblock {Searching for Primordial Black Holes in the radio and X-ray sky}.
\newblock \emph{Phys. Rev. Lett.}, 118\penalty0 (24):\penalty0 241101, 2017.
\newblock \doi{10.1103/PhysRevLett.118.241101}.

\bibitem[{Gaia Collaboration} et~al.(2016){Gaia Collaboration}, {Prusti}, {de
  Bruijne}, {Brown}, {Vallenari}, {Babusiaux}, {Bailer-Jones}, {Bastian},
  {Biermann}, {Evans}, and et~al.]{Gaia-Collaboration:2016}
{Gaia Collaboration}, T.~{Prusti}, J.~H.~J. {de Bruijne}, A.~G.~A. {Brown},
  A.~{Vallenari}, C.~{Babusiaux}, C.~A.~L. {Bailer-Jones}, U.~{Bastian},
  M.~{Biermann}, D.~W. {Evans}, and et~al.
\newblock {The Gaia mission}.
\newblock \emph{Astron.Astrophys.}, 595:\penalty0 A1, November 2016.
\newblock \doi{10.1051/0004-6361/201629272}.

\bibitem[Garcia-Bellido et~al.(2017)Garcia-Bellido, Peloso, and
  Unal]{Garcia-Bellido:2017aan}
Juan Garcia-Bellido, Marco Peloso, and Caner Unal.
\newblock {Gravitational Wave signatures of inflationary models from Primordial
  Black Hole Dark Matter}.
\newblock \emph{JCAP}, 1709\penalty0 (09):\penalty0 013, 2017.
\newblock \doi{10.1088/1475-7516/2017/09/013}.

\bibitem[Garc\'ia-Bellido et~al.(2018)Garc\'ia-Bellido, Clesse, and
  Fleury]{Garcia-Bellido:2018}
Juan Garc\'ia-Bellido, Sebastien Clesse, and Pierre Fleury.
\newblock {Primordial black holes survive SN lensing constraints}.
\newblock \emph{Phys. Dark Univ.}, 20:\penalty0 95--100, 2018.
\newblock \doi{10.1016/j.dark.2018.04.005}.

\bibitem[Gong and Gong(2018)]{Gong:2017qlj}
Yungui Gong and Yungui Gong.
\newblock {Primordial black holes and second order gravitational waves from
  ultra-slow-roll inflation}.
\newblock \emph{JCAP}, 1807\penalty0 (07):\penalty0 007, 2018.
\newblock \doi{10.1088/1475-7516/2018/07/007}.

\bibitem[{Goobar} et~al.(2009){Goobar}, {Paech}, {Stanishev}, {Amanullah},
  {Dahl{\'e}n}, {J{\"o}nsson}, {Kneib}, {Lidman}, {Limousin}, {M{\"o}rtsell},
  {Nobili}, {Richard}, {Riehm}, and {von Strauss}]{Goobar:2009}
A.~{Goobar}, K.~{Paech}, V.~{Stanishev}, R.~{Amanullah}, T.~{Dahl{\'e}n},
  J.~{J{\"o}nsson}, J.~P. {Kneib}, C.~{Lidman}, M.~{Limousin},
  E.~{M{\"o}rtsell}, S.~{Nobili}, J.~{Richard}, T.~{Riehm}, and M.~{von
  Strauss}.
\newblock {Near-IR search for lensed supernovae behind galaxy clusters. II.
  First detection and future prospects}.
\newblock \emph{\aap}, 507:\penalty0 71--83, November 2009.
\newblock \doi{10.1051/0004-6361/200811254}.

\bibitem[Green(2016)]{Green:2016}
Anne~M. Green.
\newblock {Microlensing and dynamical constraints on primordial black hole dark
  matter with an extended mass function}.
\newblock \emph{Phys. Rev.}, D94\penalty0 (6):\penalty0 063530, 2016.
\newblock \doi{10.1103/PhysRevD.94.063530}.

\bibitem[{Green} et~al.(2018){Green}, {Meerburg}, and {Meyers}]{Green:2018}
D.~{Green}, P.~D. {Meerburg}, and J.~{Meyers}.
\newblock {Aspects of Dark Matter Annihilation in Cosmology}.
\newblock \emph{ArXiv e-prints}, April 2018.

\bibitem[{Griest}(1991)]{Griest:1991}
K.~{Griest}.
\newblock {Galactic microlensing as a method of detecting massive compact halo
  objects}.
\newblock \emph{\apj}, 366:\penalty0 412--421, January 1991.
\newblock \doi{10.1086/169575}.

\bibitem[Griest et~al.(2013)Griest, Cieplak, and Lehner]{Griest:2013}
Kim Griest, Agnieszka~M. Cieplak, and Matthew~J. Lehner.
\newblock New limits on primordial black hole dark matter from an analysis of
  kepler source microlensing data.
\newblock \emph{Phys. Rev. Lett.}, 111:\penalty0 181302, Oct 2013.
\newblock \doi{10.1103/PhysRevLett.111.181302}.

\bibitem[Hawking(1975)]{Hawking:1974sw}
S.~W. Hawking.
\newblock {Particle Creation by Black Holes}.
\newblock \emph{Commun. Math. Phys.}, 43:\penalty0 199--220, 1975.
\newblock \doi{10.1007/BF02345020}.
\newblock [,167(1975)].

\bibitem[{Hawkins}(1993)]{Hawkins:1993}
M.~R.~S. {Hawkins}.
\newblock {Gravitational microlensing, quasar variability and missing matter}.
\newblock \emph{\nat}, 366:\penalty0 242--245, November 1993.
\newblock \doi{10.1038/366242a0}.

\bibitem[{Hektor} et~al.(2018){Hektor}, {H{\"u}tsi}, and {Raidal}]{Hektor:2018}
A.~{Hektor}, G.~{H{\"u}tsi}, and M.~{Raidal}.
\newblock {Constraints on primordial black hole dark matter from Galactic
  center X-ray observations}.
\newblock \emph{Astron.Astrophys.}, 618:\penalty0 A139, October 2018.
\newblock \doi{10.1051/0004-6361/201833483}.

\bibitem[Hektor et~al.(2018)Hektor, Hutsi, Marzola, Raidal, Vaskonen, and
  Veermae]{Hektor:2018qqw}
Andi Hektor, Gert Hutsi, Luca Marzola, Martti Raidal, Ville Vaskonen, and Hardi
  Veermae.
\newblock {Constraining Primordial Black Holes with the EDGES 21-cm Absorption
  Signal}.
\newblock \emph{Phys. Rev.}, D98\penalty0 (2):\penalty0 023503, 2018.
\newblock \doi{10.1103/PhysRevD.98.023503}.

\bibitem[{Helgason} et~al.(2012){Helgason}, {Ricotti}, and
  {Kashlinsky}]{Helgason:2012a}
K.~{Helgason}, M.~{Ricotti}, and A.~{Kashlinsky}.
\newblock {Reconstructing the Near-infrared Background Fluctuations from Known
  Galaxy Populations Using Multiband Measurements of Luminosity Functions}.
\newblock \emph{\apj}, 752:\penalty0 113, June 2012.
\newblock \doi{10.1088/0004-637X/752/2/113}.

\bibitem[{Helgason} et~al.(2014){Helgason}, {Cappelluti}, {Hasinger},
  {Kashlinsky}, and {Ricotti}]{Helgason:2014}
K.~{Helgason}, N.~{Cappelluti}, G.~{Hasinger}, A.~{Kashlinsky}, and
  M.~{Ricotti}.
\newblock {The Contribution of z $<$\~{} 6 Sources to the Spatial Coherence in
  the Unresolved Cosmic Near-infrared and X-Ray Backgrounds}.
\newblock \emph{\apj}, 785:\penalty0 38, April 2014.
\newblock \doi{10.1088/0004-637X/785/1/38}.

\bibitem[{Helgason} et~al.(2016){Helgason}, {Ricotti}, {Kashlinsky}, and
  {Bromm}]{Helgason:2016}
K.~{Helgason}, M.~{Ricotti}, A.~{Kashlinsky}, and V.~{Bromm}.
\newblock {On the physical requirements for a pre-reionization origin of the
  unresolved near-infrared background}.
\newblock \emph{MNRAS}, 455:\penalty0 282--294, January 2016.
\newblock \doi{10.1093/mnras/stv2209}.

\bibitem[Inomata and Nakama(2019)]{Inomata:2018epa}
Keisuke Inomata and Tomohiro Nakama.
\newblock {Gravitational waves induced by scalar perturbations as probes of the
  small-scale primordial spectrum}.
\newblock \emph{Phys. Rev.}, D99\penalty0 (4):\penalty0 043511, 2019.
\newblock \doi{10.1103/PhysRevD.99.043511}.

\bibitem[Inomata et~al.(2017)Inomata, Kawasaki, Mukaida, Tada, and
  Yanagida]{Inomata:2016rbd}
Keisuke Inomata, Masahiro Kawasaki, Kyohei Mukaida, Yuichiro Tada, and
  Tsutomu~T. Yanagida.
\newblock {Inflationary primordial black holes for the LIGO gravitational wave
  events and pulsar timing array experiments}.
\newblock \emph{Phys. Rev.}, D95\penalty0 (12):\penalty0 123510, 2017.
\newblock \doi{10.1103/PhysRevD.95.123510}.

\bibitem[Jedamzik(1997)]{Jedamzik:1996mr}
Karsten Jedamzik.
\newblock {Primordial black hole formation during the QCD epoch}.
\newblock \emph{Phys. Rev.}, D55:\penalty0 5871--5875, 1997.
\newblock \doi{10.1103/PhysRevD.55.R5871}.

\bibitem[{Kashlinsky}(2016)]{Kashlinsky:2016}
A.~{Kashlinsky}.
\newblock {LIGO Gravitational Wave Detection, Primordial Black Holes, and the
  Near-IR Cosmic Infrared Background Anisotropies}.
\newblock \emph{\apj}, 823:\penalty0 L25, June 2016.
\newblock \doi{10.3847/2041-8205/823/2/L25}.

\bibitem[{Kashlinsky} et~al.(2004){Kashlinsky}, {Arendt}, {Gardner}, {Mather},
  and {Moseley}]{Kashlinsky:2004}
A.~{Kashlinsky}, R.~{Arendt}, J.~P. {Gardner}, J.~C. {Mather}, and S.~H.
  {Moseley}.
\newblock {Detecting Population III Stars through Observations of Near-Infrared
  Cosmic Infrared Background Anisotropies}.
\newblock \emph{\apj}, 608:\penalty0 1--9, June 2004.
\newblock \doi{10.1086/386365}.

\bibitem[{Kashlinsky} et~al.(2005){Kashlinsky}, {Arendt}, {Mather}, and
  {Moseley}]{Kashlinsky:2005a}
A.~{Kashlinsky}, R.~G. {Arendt}, J.~{Mather}, and S.~H. {Moseley}.
\newblock {Tracing the first stars with fluctuations of the cosmic infrared
  background}.
\newblock \emph{\nat}, 438:\penalty0 45--50, November 2005.
\newblock \doi{10.1038/nature04143}.

\bibitem[{Kashlinsky} et~al.(2007{\natexlab{a}}){Kashlinsky}, {Arendt},
  {Mather}, and {Moseley}]{Kashlinsky:2007}
A.~{Kashlinsky}, R.~G. {Arendt}, J.~{Mather}, and S.~H. {Moseley}.
\newblock {Demonstrating the Negligible Contribution of Optical HST ACS
  Galaxies to Source-subtracted Cosmic Infrared Background Fluctuations in Deep
  Spitzer IRAC Images}.
\newblock \emph{\apjl}, 666:\penalty0 L1--L4, September 2007{\natexlab{a}}.
\newblock \doi{10.1086/521551}.

\bibitem[{Kashlinsky} et~al.(2007{\natexlab{b}}){Kashlinsky}, {Arendt},
  {Mather}, and {Moseley}]{Kashlinsky:2007a}
A.~{Kashlinsky}, R.~G. {Arendt}, J.~{Mather}, and S.~H. {Moseley}.
\newblock {New Measurements of Cosmic Infrared Background Fluctuations from
  Early Epochs}.
\newblock \emph{\apj}, 654:\penalty0 L5--L8, January 2007{\natexlab{b}}.
\newblock \doi{10.1086/510483}.

\bibitem[{Kashlinsky} et~al.(2007{\natexlab{c}}){Kashlinsky}, {Arendt},
  {Mather}, and {Moseley}]{Kashlinsky:2007b}
A.~{Kashlinsky}, R.~G. {Arendt}, J.~{Mather}, and S.~H. {Moseley}.
\newblock {On the Nature of the Sources of the Cosmic Infrared Background}.
\newblock \emph{\apj}, 654:\penalty0 L1--L4, January 2007{\natexlab{c}}.
\newblock \doi{10.1086/510484}.

\bibitem[{Kashlinsky} et~al.(2012){Kashlinsky}, {Arendt}, {Ashby}, {Fazio},
  {Mather}, and {Moseley}]{Kashlinsky:2012}
A.~{Kashlinsky}, R.~G. {Arendt}, M.~L.~N. {Ashby}, G.~G. {Fazio}, J.~{Mather},
  and S.~H. {Moseley}.
\newblock {New Measurements of the Cosmic Infrared Background Fluctuations in
  Deep Spitzer/IRAC Survey Data and Their Cosmological Implications}.
\newblock \emph{\apj}, 753:\penalty0 63, July 2012.
\newblock \doi{10.1088/0004-637X/753/1/63}.

\bibitem[{Kashlinsky} et~al.(2018){Kashlinsky}, {Arendt}, {Atrio-Barandela},
  {Cappelluti}, {Ferrara}, and {Hasinger}]{Kashlinsky:2018}
A.~{Kashlinsky}, R.~G. {Arendt}, F.~{Atrio-Barandela}, N.~{Cappelluti},
  A.~{Ferrara}, and G.~{Hasinger}.
\newblock {Looking at cosmic near-infrared background radiation anisotropies}.
\newblock \emph{Reviews of Modern Physics}, 90\penalty0 (2):\penalty0 025006,
  April 2018.
\newblock \doi{10.1103/RevModPhys.90.025006}.

\bibitem[{Kashlinsky} et~al.(2019){Kashlinsky}, {Arendt}, {Cappelluti},
  {Finoguenov}, {Hasinger}, {Helgason}, and {Merloni}]{Kashlinsky:2019}
A.~{Kashlinsky}, R.~G. {Arendt}, N.~{Cappelluti}, A.~{Finoguenov},
  G.~{Hasinger}, K.~{Helgason}, and A.~{Merloni}.
\newblock {Probing the cross-power of unresolved cosmic infrared and X-ray
  backgrounds with upcoming space missions}.
\newblock \emph{\apj}, 871:\penalty0 L6, January 2019.

\bibitem[Kessler et~al.(2019)]{Kessler:2018krb}
R.~Kessler et~al.
\newblock {First Cosmology Results using Type Ia Supernova from the Dark Energy
  Survey: Simulations to Correct Supernova Distance Biases}.
\newblock \emph{Mon. Not. Roy. Astron. Soc.}, 485:\penalty0 1171, 2019.
\newblock \doi{10.1093/mnras/stz463}.

\bibitem[Kohri et~al.(2014)Kohri, Nakama, and Suyama]{Kohri:2014lza}
Kazunori Kohri, Tomohiro Nakama, and Teruaki Suyama.
\newblock {Testing scenarios of primordial black holes being the seeds of
  supermassive black holes by ultracompact minihalos and CMB
  $\mu$-distortions}.
\newblock \emph{Phys. Rev.}, D90\penalty0 (8):\penalty0 083514, 2014.
\newblock \doi{10.1103/PhysRevD.90.083514}.

\bibitem[Koushiappas and Loeb(2017)]{Koushiappas:2017}
Savvas~M. Koushiappas and Abraham Loeb.
\newblock {Dynamics of Dwarf Galaxies Disfavor Stellar-Mass Black Holes as Dark
  Matter}.
\newblock \emph{Phys. Rev. Lett.}, 119\penalty0 (4):\penalty0 041102, 2017.
\newblock \doi{10.1103/PhysRevLett.119.041102}.

\bibitem[Lasky et~al.(2016)]{Lasky:2015lej}
Paul~D. Lasky et~al.
\newblock {Gravitational-wave cosmology across 29 decades in frequency}.
\newblock \emph{Phys. Rev.}, X6\penalty0 (1):\penalty0 011035, 2016.
\newblock \doi{10.1103/PhysRevX.6.011035}.

\bibitem[{Laureijs} et~al.(2011){Laureijs}, {Amiaux}, {Arduini},
  {Augu{\`e}res}, {Brinchmann}, {Cole}, {Cropper}, {Dabin}, {Duvet}, {Ealet},
  and et~al.]{Laureijs:2011}
R.~{Laureijs}, J.~{Amiaux}, S.~{Arduini}, J.~. {Augu{\`e}res}, J.~{Brinchmann},
  R.~{Cole}, M.~{Cropper}, C.~{Dabin}, L.~{Duvet}, A.~{Ealet}, and et~al.
\newblock {Euclid Definition Study Report}.
\newblock \emph{ArXiv e-prints}, October 2011.

\bibitem[{Laureijs} et~al.(2014){Laureijs}, {Racca}, {Stagnaro}, {Salvignol},
  {Lorenzo Alvarez}, {Saavedra Criado}, {Gaspar Venancio}, {Short}, {Strada},
  {Colombo}, {Buenadicha}, {Hoar}, {Kohley}, {Vavrek}, {Mellier}, {Berthe},
  {Amiaux}, {Cropper}, {Niemi}, {Pottinger}, {Ealet}, {Jahnke}, {Maciaszek},
  {Pasian}, {Sauvage}, {Wachter}, {Israelsson}, {Holmes}, {Seiffert},
  {Cazaubiel}, {Anselmi}, and {Musi}]{Laureijs:2014}
R.~{Laureijs}, G.~{Racca}, L.~{Stagnaro}, J.-C. {Salvignol}, J.~{Lorenzo
  Alvarez}, G.~{Saavedra Criado}, L.~{Gaspar Venancio}, A.~{Short},
  P.~{Strada}, C.~{Colombo}, G.~{Buenadicha}, J.~{Hoar}, R.~{Kohley},
  R.~{Vavrek}, Y.~{Mellier}, M.~{Berthe}, J.~{Amiaux}, M.~{Cropper},
  S.~{Niemi}, S.~{Pottinger}, A.~{Ealet}, K.~{Jahnke}, T.~{Maciaszek},
  F.~{Pasian}, M.~{Sauvage}, S.~{Wachter}, U.~{Israelsson}, W.~{Holmes},
  M.~{Seiffert}, V.~{Cazaubiel}, A.~{Anselmi}, and P.~{Musi}.
\newblock {Euclid mission status}.
\newblock In \emph{Space Telescopes and Instrumentation 2014: Optical,
  Infrared, and Millimeter Wave}, volume 9143 of \emph{Proc. SPIE}, page
  91430H, August 2014.
\newblock \doi{10.1117/12.2054883}.

\bibitem[Lazio(2013)]{Lazio:2013mea}
T.~J.~W. Lazio.
\newblock {The Square Kilometre Array pulsar timing array}.
\newblock \emph{Class. Quant. Grav.}, 30:\penalty0 224011, 2013.
\newblock \doi{10.1088/0264-9381/30/22/224011}.

\bibitem[Lentati et~al.(2015)]{Lentati:2015qwp}
L.~Lentati et~al.
\newblock {European Pulsar Timing Array Limits On An Isotropic Stochastic
  Gravitational-Wave Background}.
\newblock \emph{Mon. Not. Roy. Astron. Soc.}, 453\penalty0 (3):\penalty0
  2576--2598, 2015.
\newblock \doi{10.1093/mnras/stv1538}.

\bibitem[Li et~al.(2017)]{Li:2016}
T.~S. Li et~al.
\newblock {Farthest Neighbor: The Distant Milky Way Satellite Eridanus II}.
\newblock \emph{Astrophys. J.}, 838\penalty0 (1):\penalty0 8, 2017.
\newblock \doi{10.3847/1538-4357/aa6113}.

\bibitem[{Li} et~al.(2018){Li}, {Cappelluti}, {Arendt}, {Hasinger},
  {Kashlinsky}, and {Helgason}]{Li:2018}
Y.~{Li}, N.~{Cappelluti}, R.~G. {Arendt}, G.~{Hasinger}, A.~{Kashlinsky}, and
  K.~{Helgason}.
\newblock {The SPLASH and Chandra COSMOS Legacy Survey: The Cross-power between
  Near-infrared and X-Ray Background Fluctuations}.
\newblock \emph{\apj}, 864:\penalty0 141, September 2018.
\newblock \doi{10.3847/1538-4357/aad55a}.

\bibitem[{Lu} et~al.(2016){Lu}, {Sinukoff}, {Ofek}, {Udalski}, and
  {Kozlowski}]{Lu:2016}
J.~R. {Lu}, E.~{Sinukoff}, E.~O. {Ofek}, A.~{Udalski}, and S.~{Kozlowski}.
\newblock {A Search For Stellar-mass Black Holes Via Astrometric Microlensing}.
\newblock \emph{\apj}, 830:\penalty0 41, October 2016.
\newblock \doi{10.3847/0004-637X/830/1/41}.

\bibitem[Manchester et~al.(2013)]{Manchester:2012za}
R.~N. Manchester et~al.
\newblock {The Parkes Pulsar Timing Array Project}.
\newblock \emph{Publ. Astron. Soc. Austral.}, 30:\penalty0 17, 2013.
\newblock \doi{10.1017/pasa.2012.017}.

\bibitem[Mandic et~al.(2016)Mandic, Bird, and Cholis]{Mandic:2016lcn}
Vuk Mandic, Simeon Bird, and Ilias Cholis.
\newblock {Stochastic Gravitational-Wave Background due to Primordial Binary
  Black Hole Mergers}.
\newblock \emph{Phys. Rev. Lett.}, 117\penalty0 (20):\penalty0 201102, 2016.
\newblock \doi{10.1103/PhysRevLett.117.201102}.

\bibitem[{Mediavilla} et~al.(2017){Mediavilla}, {Jim{\'e}nez-Vicente},
  {Mu{\~n}oz}, {Vives-Arias}, and {Calder{\'o}n-Infante}]{Mediavilla:2017}
E.~{Mediavilla}, J.~{Jim{\'e}nez-Vicente}, J.~A. {Mu{\~n}oz}, H.~{Vives-Arias},
  and J.~{Calder{\'o}n-Infante}.
\newblock {Limits on the Mass and Abundance of Primordial Black Holes from
  Quasar Gravitational Microlensing}.
\newblock \emph{\apj}, 836:\penalty0 L18, February 2017.
\newblock \doi{10.3847/2041-8213/aa5dab}.

\bibitem[{Merloni} et~al.(2012){Merloni}, {Predehl}, {Becker}, {B{\"o}hringer},
  {Boller}, {Brunner}, {Brusa}, {Dennerl}, {Freyberg}, {Friedrich},
  {Georgakakis}, {Haberl}, {Hasinger}, {Meidinger}, {Mohr}, {Nandra}, {Rau},
  {Reiprich}, {Robrade}, {Salvato}, {Santangelo}, {Sasaki}, {Schwope}, {Wilms},
  and {German eROSITA Consortium}]{Merloni:2012}
A.~{Merloni}, P.~{Predehl}, W.~{Becker}, H.~{B{\"o}hringer}, T.~{Boller},
  H.~{Brunner}, M.~{Brusa}, K.~{Dennerl}, M.~{Freyberg}, P.~{Friedrich},
  A.~{Georgakakis}, F.~{Haberl}, G.~{Hasinger}, N.~{Meidinger}, J.~{Mohr},
  K.~{Nandra}, A.~{Rau}, T.~H. {Reiprich}, J.~{Robrade}, M.~{Salvato},
  A.~{Santangelo}, M.~{Sasaki}, A.~{Schwope}, J.~{Wilms}, and t.~{German
  eROSITA Consortium}.
\newblock {eROSITA Science Book: Mapping the Structure of the Energetic
  Universe}.
\newblock \emph{ArXiv e-prints}, September 2012.

\bibitem[{Mesinger} et~al.(2006){Mesinger}, {Johnson}, and
  {Haiman}]{Mesinger:2006}
A.~{Mesinger}, B.~D. {Johnson}, and Z.~{Haiman}.
\newblock {The Redshift Distribution of Distant Supernovae and Its Use in
  Probing Reionization}.
\newblock \emph{\apj}, 637:\penalty0 80--90, January 2006.
\newblock \doi{10.1086/498294}.

\bibitem[{Meszaros}(1974)]{Meszaros:1974}
P.~{Meszaros}.
\newblock {The behaviour of point masses in an expanding cosmological
  substratum}.
\newblock \emph{Astron.Astrophys.}, 37:\penalty0 225--228, December 1974.

\bibitem[Metcalf and Silk(1999)]{Metcalf:1999qb}
R.~Benton Metcalf and Joseph Silk.
\newblock {A Fundamental test of the nature of dark matter}.
\newblock \emph{Astrophys. J.}, 519:\penalty0 L1--4, 1999.
\newblock \doi{10.1086/312086}.

\bibitem[{Mitchell-Wynne} et~al.(2016){Mitchell-Wynne}, {Cooray}, {Xue}, {Luo},
  {Brandt}, and {Koekemoer}]{Mitchell-Wynne:2016}
K.~{Mitchell-Wynne}, A.~{Cooray}, Y.~{Xue}, B.~{Luo}, W.~{Brandt}, and
  A.~{Koekemoer}.
\newblock {Cross-correlation between X-Ray and Optical/Near-infrared Background
  Intensity Fluctuations}.
\newblock \emph{\apj}, 832:\penalty0 104, December 2016.
\newblock \doi{10.3847/0004-637X/832/2/104}.

\bibitem[{Monroy-Rodr{\'{\i}}guez} and {Allen}(2014)]{Monroy-Rodriguez:2014}
M.~A. {Monroy-Rodr{\'{\i}}guez} and C.~{Allen}.
\newblock {The End of the MACHO Era, Revisited: New Limits on MACHO Masses from
  Halo Wide Binaries}.
\newblock \emph{Astrophys. J.}, 790:\penalty0 159, August 2014.
\newblock \doi{10.1088/0004-637X/790/2/159}.

\bibitem[{Mu{\~n}oz} et~al.(2016){Mu{\~n}oz}, {Kovetz}, {Dai}, and
  {Kamionkowski}]{Munoz:2016}
J.~B. {Mu{\~n}oz}, E.~D. {Kovetz}, L.~{Dai}, and M.~{Kamionkowski}.
\newblock {Lensing of Fast Radio Bursts as a Probe of Compact Dark Matter}.
\newblock \emph{Physical Review Letters}, 117\penalty0 (9):\penalty0 091301,
  August 2016.
\newblock \doi{10.1103/PhysRevLett.117.091301}.

\bibitem[Nakama et~al.(2017)Nakama, Silk, and Kamionkowski]{Nakama:2016gzw}
Tomohiro Nakama, Joseph Silk, and Marc Kamionkowski.
\newblock {Stochastic gravitational waves associated with the formation of
  primordial black holes}.
\newblock \emph{Phys. Rev.}, D95\penalty0 (4):\penalty0 043511, 2017.
\newblock \doi{10.1103/PhysRevD.95.043511}.

\bibitem[{Nakamura} et~al.(1997){Nakamura}, {Sasaki}, {Tanaka}, and
  {Thorne}]{Nakamura_97}
T.~{Nakamura}, M.~{Sasaki}, T.~{Tanaka}, and K.~S. {Thorne}.
\newblock {Gravitational Waves from Coalescing Black Hole MACHO Binaries}.
\newblock \emph{\apjl}, 487:\penalty0 L139--L142, October 1997.
\newblock \doi{10.1086/310886}.

\bibitem[{Nandra} et~al.(2013){Nandra}, {Barret}, {Barcons}, {Fabian}, {den
  Herder}, {Piro}, {Watson}, {Adami}, {Aird}, {Afonso}, and
  et~al.]{Nandra:2013}
K.~{Nandra}, D.~{Barret}, X.~{Barcons}, A.~{Fabian}, J.-W. {den Herder},
  L.~{Piro}, M.~{Watson}, C.~{Adami}, J.~{Aird}, J.~M. {Afonso}, and et~al.
\newblock {The Hot and Energetic Universe: A White Paper presenting the science
  theme motivating the Athena+ mission}.
\newblock \emph{ArXiv e-prints}, June 2013.

\bibitem[{Navarro} et~al.(2018){Navarro}, {Minniti}, and
  {Contreras-Ramos}]{Navarro2018}
M.~G. {Navarro}, D.~{Minniti}, and R.~{Contreras-Ramos}.
\newblock {VVV Survey Microlensing: The Galactic Longitude Dependence}.
\newblock \emph{ApJL}, 865:\penalty0 L5, September 2018.
\newblock \doi{10.3847/2041-8213/aae08a}.

\bibitem[Orlofsky et~al.(2017)Orlofsky, Pierce, and Wells]{Orlofsky:2016vbd}
Nicholas Orlofsky, Aaron Pierce, and James~D. Wells.
\newblock {Inflationary theory and pulsar timing investigations of primordial
  black holes and gravitational waves}.
\newblock \emph{Phys. Rev.}, D95\penalty0 (6):\penalty0 063518, 2017.
\newblock \doi{10.1103/PhysRevD.95.063518}.

\bibitem[{Paczynski}(1986)]{Paczynski:1986}
B.~{Paczynski}.
\newblock {Gravitational microlensing by the galactic halo}.
\newblock \emph{\apj}, 304:\penalty0 1--5, May 1986.
\newblock \doi{10.1086/164140}.

\bibitem[Pani and Loeb(2013)]{Pani:2013hpa}
Paolo Pani and Abraham Loeb.
\newblock {Constraining Primordial Black-Hole Bombs through Spectral
  Distortions of the Cosmic Microwave Background}.
\newblock \emph{Phys. Rev.}, D88:\penalty0 041301, 2013.
\newblock \doi{10.1103/PhysRevD.88.041301}.

\bibitem[{Poulin} et~al.(2017){Poulin}, {Serpico}, {Calore}, {Clesse}, and
  {Kohri}]{Poulin:2017}
V.~{Poulin}, P.~D. {Serpico}, F.~{Calore}, S.~{Clesse}, and K.~{Kohri}.
\newblock {CMB bounds on disk-accreting massive primordial black holes}.
\newblock \emph{\prd}, 96\penalty0 (8):\penalty0 083524, October 2017.
\newblock \doi{10.1103/PhysRevD.96.083524}.

\bibitem[Raidal et~al.(2017)Raidal, Vaskonen, and Veerm{\"a}e]{Raidal:2017mfl}
Martti Raidal, Ville Vaskonen, and Hardi Veerm{\"a}e.
\newblock {Gravitational Waves from Primordial Black Hole Mergers}.
\newblock \emph{JCAP}, 1709:\penalty0 037, 2017.
\newblock \doi{10.1088/1475-7516/2017/09/037}.

\bibitem[Raidal et~al.(2019)Raidal, Spethmann, Vaskonen, and
  Veermäe]{Raidal_18}
Martti Raidal, Christian Spethmann, Ville Vaskonen, and Hardi Veermäe.
\newblock {Formation and Evolution of Primordial Black Hole Binaries in the
  Early Universe}.
\newblock \emph{JCAP}, 1902:\penalty0 018, 2019.
\newblock \doi{10.1088/1475-7516/2019/02/018}.

\bibitem[{Rauch}(1991)]{Rauch:1991}
K.~P. {Rauch}.
\newblock {Gravitational microlensing of high-redshift supernovae by compact
  objects}.
\newblock \emph{\apj}, 374:\penalty0 83--90, June 1991.
\newblock \doi{10.1086/170098}.

\bibitem[{Ricotti} et~al.(2008){Ricotti}, {Ostriker}, and {Mack}]{Ricotti:2008}
M.~{Ricotti}, J.~P. {Ostriker}, and K.~J. {Mack}.
\newblock {Effect of Primordial Black Holes on the Cosmic Microwave Background
  and Cosmological Parameter Estimates}.
\newblock \emph{\apj}, 680:\penalty0 829--845, June 2008.
\newblock \doi{10.1086/587831}.

\bibitem[{Rodney} et~al.(2014){Rodney}, {Riess}, {Strolger}, {Dahlen}, {Graur},
  {Casertano}, {Dickinson}, {Ferguson}, {Garnavich}, {Hayden}, {Jha}, {Jones},
  {Kirshner}, {Koekemoer}, {McCully}, {Mobasher}, {Patel}, {Weiner}, {Cenko},
  {Clubb}, {Cooper}, {Filippenko}, {Frederiksen}, {Hjorth}, {Leibundgut},
  {Matheson}, {Nayyeri}, {Penner}, {Trump}, {Silverman}, {U}, {Azalee
  Bostroem}, {Challis}, {Rajan}, {Wolff}, {Faber}, {Grogin}, and
  {Kocevski}]{Rodney:2014}
S.~A. {Rodney}, A.~G. {Riess}, L.-G. {Strolger}, T.~{Dahlen}, O.~{Graur},
  S.~{Casertano}, M.~E. {Dickinson}, H.~C. {Ferguson}, P.~{Garnavich},
  B.~{Hayden}, S.~W. {Jha}, D.~O. {Jones}, R.~P. {Kirshner}, A.~M. {Koekemoer},
  C.~{McCully}, B.~{Mobasher}, B.~{Patel}, B.~J. {Weiner}, S.~B. {Cenko}, K.~I.
  {Clubb}, M.~{Cooper}, A.~V. {Filippenko}, T.~F. {Frederiksen}, J.~{Hjorth},
  B.~{Leibundgut}, T.~{Matheson}, H.~{Nayyeri}, K.~{Penner}, J.~{Trump}, J.~M.
  {Silverman}, V.~{U}, K.~{Azalee Bostroem}, P.~{Challis}, A.~{Rajan},
  S.~{Wolff}, S.~M. {Faber}, N.~A. {Grogin}, and D.~{Kocevski}.
\newblock {Type Ia Supernova Rate Measurements to Redshift 2.5 from CANDELS:
  Searching for Prompt Explosions in the Early Universe}.
\newblock \emph{\aj}, 148:\penalty0 13, July 2014.
\newblock \doi{10.1088/0004-6256/148/1/13}.

\bibitem[{Rybicki} et~al.(2018){Rybicki}, {Wyrzykowski}, {Klencki}, {de
  Bruijne}, {Belczy{\'n}ski}, and {Chru{\'s}li{\'n}ska}]{Rybicki:2018}
K.~A. {Rybicki}, {\L}.~{Wyrzykowski}, J.~{Klencki}, J.~{de Bruijne},
  K.~{Belczy{\'n}ski}, and M.~{Chru{\'s}li{\'n}ska}.
\newblock {On the accuracy of mass measurement for microlensing black holes as
  seen by Gaia and OGLE}.
\newblock \emph{MNRAS}, 476:\penalty0 2013--2028, May 2018.
\newblock \doi{10.1093/mnras/sty356}.

\bibitem[{Sajadian} and {Poleski}(2018)]{SajadianPoleski:2018}
S.~{Sajadian} and R.~{Poleski}.
\newblock {Prediction on detection and characterization of Galactic disk
  microlensing events by LSST}.
\newblock \emph{ArXiv e-prints}, June 2018.

\bibitem[{Sasaki} et~al.(2016){Sasaki}, {Suyama}, {Tanaka}, and
  {Yokoyama}]{Sasaki_16}
M.~{Sasaki}, T.~{Suyama}, T.~{Tanaka}, and S.~{Yokoyama}.
\newblock {Primordial Black Hole Scenario for the Gravitational-Wave Event
  GW150914}.
\newblock \emph{Physical Review Letters}, 117\penalty0 (6):\penalty0 061101,
  August 2016.
\newblock \doi{10.1103/PhysRevLett.117.061101}.

\bibitem[Seljak and Holz(1999)]{Seljak:1999tm}
Uros Seljak and Daniel~E. Holz.
\newblock {Limits on the density of compact objects from high redshift
  supernovae}.
\newblock \emph{Astron. Astrophys.}, 351:\penalty0 L10, 1999.

\bibitem[{Spergel} et~al.(2015){Spergel}, {Gehrels}, {Baltay}, { },
  {Breckinridge}, {Donahue}, {Dressler}, {Gaudi}, {Greene}, {Guyon}, {Hirata},
  {Kalirai}, {Kasdin}, {Macintosh}, {Moos}, {Perlmutter}, {Postman},
  {Rauscher}, {Rhodes}, {Wang}, {Weinberg}, {Benford}, {Hudson}, {Jeong},
  {Mellier}, {Traub}, {Yamada}, {Capak}, {Colbert}, {Masters}, {Penny},
  {Savransky}, {Stern}, {Zimmerman}, {Barry}, {Bartusek}, {Carpenter}, {Cheng},
  {Content}, {Dekens}, {Demers}, {Grady}, {Jackson}, {Kuan}, {Kruk}, {Melton},
  {Nemati}, {Parvin}, {Poberezhskiy}, {Peddie}, {Ruffa}, {Wallace}, {Whipple},
  {Wollack}, and {Zhao}]{Spergel:2015}
D.~{Spergel}, N.~{Gehrels}, C.~{Baltay}, D.~{ }, J.~{Breckinridge},
  M.~{Donahue}, A.~{Dressler}, B.~S. {Gaudi}, T.~{Greene}, O.~{Guyon},
  C.~{Hirata}, J.~{Kalirai}, N.~J. {Kasdin}, B.~{Macintosh}, W.~{Moos},
  S.~{Perlmutter}, M.~{Postman}, B.~{Rauscher}, J.~{Rhodes}, Y.~{Wang},
  D.~{Weinberg}, D.~{Benford}, M.~{Hudson}, W.-S. {Jeong}, Y.~{Mellier},
  W.~{Traub}, T.~{Yamada}, P.~{Capak}, J.~{Colbert}, D.~{Masters}, M.~{Penny},
  D.~{Savransky}, D.~{Stern}, N.~{Zimmerman}, R.~{Barry}, L.~{Bartusek},
  K.~{Carpenter}, E.~{Cheng}, D.~{Content}, F.~{Dekens}, R.~{Demers},
  K.~{Grady}, C.~{Jackson}, G.~{Kuan}, J.~{Kruk}, M.~{Melton}, B.~{Nemati},
  B.~{Parvin}, I.~{Poberezhskiy}, C.~{Peddie}, J.~{Ruffa}, J.~K. {Wallace},
  A.~{Whipple}, E.~{Wollack}, and F.~{Zhao}.
\newblock {Wide-Field InfrarRed Survey Telescope-Astrophysics Focused Telescope
  Assets WFIRST-AFTA 2015 Report}.
\newblock \emph{ArXiv e-prints}, March 2015.

\bibitem[{Strolger} et~al.(2004){Strolger}, {Riess}, {Dahlen}, {Livio},
  {Panagia}, {Challis}, {Tonry}, {Filippenko}, {Chornock}, {Ferguson},
  {Koekemoer}, {Mobasher}, {Dickinson}, {Giavalisco}, {Casertano}, {Hook},
  {Blondin}, {Leibundgut}, {Nonino}, {Rosati}, {Spinrad}, {Steidel}, {Stern},
  {Garnavich}, {Matheson}, {Grogin}, {Hornschemeier}, {Kretchmer}, {Laidler},
  {Lee}, {Lucas}, {de Mello}, {Moustakas}, {Ravindranath}, {Richardson}, and
  {Taylor}]{Strolger:2004}
L.-G. {Strolger}, A.~G. {Riess}, T.~{Dahlen}, M.~{Livio}, N.~{Panagia},
  P.~{Challis}, J.~L. {Tonry}, A.~V. {Filippenko}, R.~{Chornock},
  H.~{Ferguson}, A.~{Koekemoer}, B.~{Mobasher}, M.~{Dickinson},
  M.~{Giavalisco}, S.~{Casertano}, R.~{Hook}, S.~{Blondin}, B.~{Leibundgut},
  M.~{Nonino}, P.~{Rosati}, H.~{Spinrad}, C.~C. {Steidel}, D.~{Stern}, P.~M.
  {Garnavich}, T.~{Matheson}, N.~{Grogin}, A.~{Hornschemeier}, C.~{Kretchmer},
  V.~G. {Laidler}, K.~{Lee}, R.~{Lucas}, D.~{de Mello}, L.~A. {Moustakas},
  S.~{Ravindranath}, M.~{Richardson}, and E.~{Taylor}.
\newblock {The Hubble Higher z Supernova Search: Supernovae to z \~{} 1.6 and
  Constraints on Type Ia Progenitor Models}.
\newblock \emph{\apj}, 613:\penalty0 200--223, September 2004.
\newblock \doi{10.1086/422901}.

\bibitem[Telescope()]{site:ET}
Einstein Telescope.
\newblock URL \url{http://www.et-gw.eu/}.

\bibitem[{The LIGO Scientific Collaboration} et~al.(2019){The LIGO Scientific
  Collaboration}, {the Virgo Collaboration}, {Abbott},
  et~al.]{The-LIGO-Scientific-Collaboration:2019}
{The LIGO Scientific Collaboration}, {the Virgo Collaboration}, B.~P. {Abbott},
  et~al.
\newblock {Low-Latency Gravitational Wave Alerts for Multi-Messenger Astronomy
  During the Second Advanced LIGO and Virgo Observing Run}.
\newblock \emph{arXiv e-prints}, art. arXiv:1901.03310, Jan 2019.

\bibitem[{The Simons Observatory Collaboration} et~al.(2018){The Simons
  Observatory Collaboration}, {Ade}, {Aguirre}, {Ahmed}, {Aiola}, {Ali},
  {Alonso}, {Alvarez}, {Arnold}, {Ashton}, and
  et~al.]{The-Simons-Observatory-Collaboration:2018}
{The Simons Observatory Collaboration}, P.~{Ade}, J.~{Aguirre}, Z.~{Ahmed},
  S.~{Aiola}, A.~{Ali}, D.~{Alonso}, M.~A. {Alvarez}, K.~{Arnold}, P.~{Ashton},
  and et~al.
\newblock {The Simons Observatory: Science goals and forecasts}.
\newblock \emph{arXiv e-prints}, August 2018.

\bibitem[{Tisserand} et~al.(2007){Tisserand}, {Le Guillou}, {Afonso}, {Albert},
  {Andersen}, {Ansari}, {Aubourg}, {Bareyre}, {Beaulieu}, {Charlot},
  {Coutures}, {Ferlet}, {Fouqu{\'e}}, {Glicenstein}, {Goldman}, {Gould},
  {Graff}, {Gros}, {Haissinski}, {Hamadache}, {de Kat}, {Lasserre}, {Lesquoy},
  {Loup}, {Magneville}, {Marquette}, {Maurice}, {Maury}, {Milsztajn}, {Moniez},
  {Palanque-Delabrouille}, {Perdereau}, {Rahal}, {Rich}, {Spiro},
  {Vidal-Madjar}, {Vigroux}, {Zylberajch}, and {EROS-2
  Collaboration}]{Tisserand:2007}
P.~{Tisserand}, L.~{Le Guillou}, C.~{Afonso}, J.~N. {Albert}, J.~{Andersen},
  R.~{Ansari}, {\'E}.~{Aubourg}, P.~{Bareyre}, J.~P. {Beaulieu}, X.~{Charlot},
  C.~{Coutures}, R.~{Ferlet}, P.~{Fouqu{\'e}}, J.~F. {Glicenstein},
  B.~{Goldman}, A.~{Gould}, D.~{Graff}, M.~{Gros}, J.~{Haissinski},
  C.~{Hamadache}, J.~{de Kat}, T.~{Lasserre}, {\'E}.~{Lesquoy}, C.~{Loup},
  C.~{Magneville}, J.~B. {Marquette}, {\'E}.~{Maurice}, A.~{Maury},
  A.~{Milsztajn}, M.~{Moniez}, N.~{Palanque-Delabrouille}, O.~{Perdereau},
  Y.~R. {Rahal}, J.~{Rich}, M.~{Spiro}, A.~{Vidal-Madjar}, L.~{Vigroux},
  S.~{Zylberajch}, and {EROS-2 Collaboration}.
\newblock {Limits on the Macho content of the Galactic Halo from the EROS-2
  Survey of the Magellanic Clouds}.
\newblock \emph{Astron.Astrophys.}, 469:\penalty0 387--404, July 2007.
\newblock \doi{10.1051/0004-6361:20066017}.

\bibitem[{Tonry} et~al.(2003){Tonry}, {Schmidt}, {Barris}, {Candia}, {Challis},
  {Clocchiatti}, {Coil}, {Filippenko}, {Garnavich}, {Hogan}, {Holland}, {Jha},
  {Kirshner}, {Krisciunas}, {Leibundgut}, {Li}, {Matheson}, {Phillips},
  {Riess}, {Schommer}, {Smith}, {Sollerman}, {Spyromilio}, {Stubbs}, and
  {Suntzeff}]{Tonry:2003}
J.~L. {Tonry}, B.~P. {Schmidt}, B.~{Barris}, P.~{Candia}, P.~{Challis},
  A.~{Clocchiatti}, A.~L. {Coil}, A.~V. {Filippenko}, P.~{Garnavich},
  C.~{Hogan}, S.~T. {Holland}, S.~{Jha}, R.~P. {Kirshner}, K.~{Krisciunas},
  B.~{Leibundgut}, W.~{Li}, T.~{Matheson}, M.~M. {Phillips}, A.~G. {Riess},
  R.~{Schommer}, R.~C. {Smith}, J.~{Sollerman}, J.~{Spyromilio}, C.~W.
  {Stubbs}, and N.~B. {Suntzeff}.
\newblock {Cosmological Results from High-z Supernovae}.
\newblock \emph{\apj}, 594:\penalty0 1--24, September 2003.
\newblock \doi{10.1086/376865}.

\bibitem[Verbiest et~al.(2016)]{Verbiest:2016vem}
J.~P.~W. Verbiest et~al.
\newblock {The International Pulsar Timing Array: First Data Release}.
\newblock \emph{Mon. Not. Roy. Astron. Soc.}, 458\penalty0 (2):\penalty0
  1267--1288, 2016.
\newblock \doi{10.1093/mnras/stw347}.

\bibitem[{Wyrzykowski} et~al.(2009){Wyrzykowski}, {Koz{\l}owski}, {Skowron},
  {Belokurov}, {Smith}, {Udalski}, {Szyma{\'n}ski}, {Kubiak},
  {Pietrzy{\'n}ski}, {Soszy{\'n}ski}, {Szewczyk}, and
  {{\.Z}ebru{\'n}}]{Wyrzykowski:2009}
{\L}.~{Wyrzykowski}, S.~{Koz{\l}owski}, J.~{Skowron}, V.~{Belokurov}, M.~C.
  {Smith}, A.~{Udalski}, M.~K. {Szyma{\'n}ski}, M.~{Kubiak},
  G.~{Pietrzy{\'n}ski}, I.~{Soszy{\'n}ski}, O.~{Szewczyk}, and
  K.~{{\.Z}ebru{\'n}}.
\newblock {The OGLE view of microlensing towards the Magellanic Clouds - I. A
  trickle of events in the OGLE-II LMC data}.
\newblock \emph{MNRAS}, 397:\penalty0 1228--1242, August 2009.
\newblock \doi{10.1111/j.1365-2966.2009.15029.x}.

\bibitem[{Wyrzykowski} et~al.(2011){Wyrzykowski}, {Koz{\l}owski}, {Skowron},
  {Udalski}, {Szyma{\'n}ski}, {Kubiak}, {Pietrzy{\'n}ski}, {Soszy{\'n}ski},
  {Szewczyk}, {Ulaczyk}, and {Poleski}]{Wyrzykowski:2011a}
{\L}.~{Wyrzykowski}, S.~{Koz{\l}owski}, J.~{Skowron}, A.~{Udalski}, M.~K.
  {Szyma{\'n}ski}, M.~{Kubiak}, G.~{Pietrzy{\'n}ski}, I.~{Soszy{\'n}ski},
  O.~{Szewczyk}, K.~{Ulaczyk}, and R.~{Poleski}.
\newblock {The OGLE view of microlensing towards the Magellanic Clouds - III.
  Ruling out subsolar MACHOs with the OGLE-III LMC data}.
\newblock \emph{MNRAS}, 413:\penalty0 493--508, May 2011.
\newblock \doi{10.1111/j.1365-2966.2010.18150.x}.

\bibitem[{Wyrzykowski} et~al.(2016){Wyrzykowski}, {Kostrzewa-Rutkowska},
  {Skowron}, {Rybicki}, {Mr{\'o}z}, {Koz{\l}owski}, {Udalski}, {Szyma{\'n}ski},
  {Pietrzy{\'n}ski}, {Soszy{\'n}ski}, {Ulaczyk}, {Pietrukowicz}, {Poleski},
  {Pawlak}, {I{\l}kiewicz}, and {Rattenbury}]{Wyrzykowski:2016}
{\L}.~{Wyrzykowski}, Z.~{Kostrzewa-Rutkowska}, J.~{Skowron}, K.~A. {Rybicki},
  P.~{Mr{\'o}z}, S.~{Koz{\l}owski}, A.~{Udalski}, M.~K. {Szyma{\'n}ski},
  G.~{Pietrzy{\'n}ski}, I.~{Soszy{\'n}ski}, K.~{Ulaczyk}, P.~{Pietrukowicz},
  R.~{Poleski}, M.~{Pawlak}, K.~{I{\l}kiewicz}, and N.~J. {Rattenbury}.
\newblock {Black hole, neutron star and white dwarf candidates from
  microlensing with OGLE-III}.
\newblock \emph{MNRAS}, 458:\penalty0 3012--3026, May 2016.
\newblock \doi{10.1093/mnras/stw426}.

\bibitem[Zhao et~al.(2013)Zhao, Zhang, You, and Zhu]{Zhao:2013bba}
Wen Zhao, Yang Zhang, Xiao-Peng You, and Zong-Hong Zhu.
\newblock {Constraints of relic gravitational waves by pulsar timing arrays:
  Forecasts for the FAST and SKA projects}.
\newblock \emph{Phys. Rev.}, D87\penalty0 (12):\penalty0 124012, 2013.
\newblock \doi{10.1103/PhysRevD.87.124012}.

\bibitem[Zumalacarregui and Seljak(2018)]{Zumalacarregui:2018}
Miguel Zumalacarregui and Uros Seljak.
\newblock {Limits on stellar-mass compact objects as dark matter from
  gravitational lensing of type Ia supernovae}.
\newblock \emph{Phys. Rev. Lett.}, 121\penalty0 (14):\penalty0 141101, 2018.
\newblock \doi{10.1103/PhysRevLett.121.141101}.

\end{thebibliography}

\end{document}